\documentclass[twocolumn,trackchanges]{aastex701}
\usepackage{graphicx}
\usepackage{tabularx}
\usepackage{appendix}
\usepackage{siunitx}
\usepackage{amsmath}
\usepackage{txfonts}
\usepackage{xcolor}
\usepackage{soul}
\usepackage{caption}
\usepackage{comment}
\usepackage{subcaption}
\usepackage{booktabs}
\usepackage{enumitem}
\usepackage{natbib}
\usepackage{csquotes}
\usepackage{float}
\usepackage{makecell}
\usepackage{multirow}
\usepackage{xspace}

\newcommand{\bpop}{\textsc{B-pop~}}
\newcommand{\dragon}{\textsc{DRAGON-II}}
\newcommand{\globular}{GCs\xspace}
\newcommand{\nuclear}{NCs\xspace}
\newcommand{\young}{YCs\xspace}
\newcommand{\msun}{{\,\rm M}_\odot}

\received{\today}
\revised{\today}
\accepted{\today}
\submitjournal{ApJL}
\graphicspath{{./}{./}}

\begin{document}


\title{Assembling GW231123 in star clusters through the combination of stellar binary evolution and hierarchical mergers}

\author[orcid=0009-0001-7605-991X,sname='Lavinia Paiella']{Lavinia Paiella}
\affiliation{Gran Sasso Science Institute (GSSI), L'Aquila (Italy), Viale Francesco Crispi 7}
\affiliation{INFN, Laboratori Nazionali del Gran Sasso, 67100 Assergi, Italy}
\email[show]{lavinia.paiella@gssi.it\newline}  

\author[orcid=0009-0005-9890-4722,sname='Cristiano Ugolini']{Cristiano Ugolini}
\affiliation{Gran Sasso Science Institute (GSSI), L'Aquila (Italy), Viale Francesco Crispi 7}
\affiliation{INFN, Laboratori Nazionali del Gran Sasso, 67100 Assergi, Italy}
\email[show]{cristiano.ugolini@gssi.it\newline} 

\author[orcid=0000-0003-0930-6930, sname='Mario Spera']{Mario Spera}
\affiliation{
International School for Advanced Studies (SISSA), Via Bonomea 265, I-34136 Trieste, Italy}
\affiliation{National Institute for Nuclear Physics - INFN, Sezione di Trieste, I-34127 Trieste, Italy}
\affiliation{Istituto Nazionale di Astrofisica - Osservatorio Astronomico di Roma, Via Frascati 33, I-00040, Monteporzio Catone, Italy}
\email[]{}

\author[orcid=0000-0003-1643-0526,sname='Marica Branchesi']{Marica Branchesi}
\affiliation{Gran Sasso Science Institute (GSSI), L'Aquila (Italy), Viale Francesco Crispi 7}
\affiliation{INFN, Laboratori Nazionali del Gran Sasso, 67100 Assergi, Italy}
\email[]{}

\author[orcid=0000-0002-3987-0519,sname='Manuel Arca Sedda']{Manuel Arca Sedda}
\affiliation{Gran Sasso Science Institute (GSSI), L'Aquila (Italy), Viale Francesco Crispi 7}
\affiliation{INFN, Laboratori Nazionali del Gran Sasso, 67100 Assergi, Italy}
\email[show]{manuel.arcasedda@gssi.it}  

\begin{abstract}
GW231123 is the most massive binary black hole (BBH) merger detected to date by the LIGO-Virgo-KAGRA collaboration. With at least one black hole (BH) in the upper-mass gap and both BHs exhibiting high spins ($\chi_{1,2} \gtrsim 0.8$), this event challenges standard isolated binary evolution models. A compelling alternative is a dynamical origin in star clusters, where stellar binaries and hierarchical mergers may both contribute to the formation of similar BBHs. In this work, we investigate the formation of GW231123-like events in different cluster environments using the \textsc{B-pop~}semi-analytic population synthesis code. We find that low-metallicity environments ($Z \lesssim 0.002$) are ideal for producing BBH mergers similar to GW231123. In young and globular clusters, these BBHs have components formed in stellar binaries, whilst in nuclear clusters there is also a significant contribution from BHs built-up via hierarchical mergers.
Natal spins of BHs formed in stellar binaries are crucial to find GW231123 analogs. In particular, our models suggest that BHs from stellar binaries are likely characterized by high-spins. Simulated GW231123-like systems exhibit short delay times, $t_\mathrm{del} \sim 0.1 - 1$ Gyr, which suggests their progenitors formed close to the inferred merger redshift ($z = 0.39^{+0.27}_{-0.24}$). We argue that star clusters in metal-poor dwarf galaxies or Milky Way-like galaxies are ideal nurseries, inferring an upper limit to the local merger rate of $\mathcal{R} \sim 1.6\times10^{-3} - 0.16$ yr$^{-1}$ Gpc$^{-3}$ for nuclear clusters, $\sim 0.036 - 0.72$ yr$^{-1}$ Gpc$^{-3}$ for globular clusters, and $4\times10^{-4}-0.041$ yr$^{-1}$ Gpc$^{-3}$ for young clusters.
\end{abstract}

\keywords{\uat{Astrophysical black holes}{98} --- \uat{Stellar dynamics}{1596} --- \uat{Gravitational waves}{678}}

\section{Introduction}

On 2023 November 23, the LIGO-Virgo-KAGRA (LVK) collaboration detected GW231123, a gravitational-wave (GW) source associated with the merger of two black holes (BHs) with component masses of $m_1 = 137^{+22}_{-17}\msun$, and $m_2 = 103^{+20}_{-52}\msun$ \citep{GW231123_paper}. The inferred masses make GW231123 the most massive binary black hole (BBH) merger ever detected, and possibly the first one involving two intermediate-mass black holes (IMBHs). Additionally, data analysis of the event suggests that both BHs are highly spinning (with median spin values around $\chi_{1,2} \simeq 0.8$).
From the astrophysical viewpoint, GW231123 challenges current stellar evolution theories. Both BHs masses potentially fall in the upper-mass gap, a range extending from $60-80\msun$ up to $\sim 220\msun$ \citep[see e.g.][]{renzoPairinstabilityEvolutionExplosions2024}, where the formation of BHs from stellar collapse is severely suppressed by the onset of pair-instability (PISN) and pulsational pair-instability supernova \citep[PPISN,][]{Woosley2017, speraVeryMassiveStars2017, iorioCompactObjectMergers2023, Ugolini2025}. These explosive processes shape the mass spectrum of merging BBHs formed from the isolated evolution of binary stars, leading to mergers sensibly lighter than GW231123 \citep[see e.g.][]{speraMergingBlackHole2019a, mapelliFormationChannelsSingle2021a, vansonFillingGapPossible2021,iorioCompactObjectMergers2023, hendriksPulsationalPairinstabilitySupernovae2023}. 

Dynamical interactions in star clusters represents another viable pathway to the formation of merging BBHs \citep[see e.g.][]{Heggie1975,Miller_IMBH_2002,miller_four-body_2002, Portegies_Zwart2006, banerjeeStellarmassBlackHoles2010,Banerjee2018, rodriguezBinaryBlackHole2015, mapelliCosmicEvolutionBinary2022a,dicarloMergingBlackHoles2019a, rodriguezBlackHolesNext2019, 2000A&A...359..695P, Antonini_2019, arca_sedda_fingerprints_2020,AS_DragonII_3}.
The formation of BHs with masses comparable to those inferred from GW231123 signal requires either multiple collisions among stars, or repeated --- hierarchical --- BBH mergers. 
Stellar collision products, for instance, can reach a chemical composition and structure such to avoid (P)PISN and undergo direct collapse, forming BHs in the upper-mass gap and beyond \citep[]{Kremer2020, DiCarlo2021a, Costa2022,Ballone2023, Arca_Sedda_2023}. While stellar collisions and mergers can also occur in isolated binaries, depending on the orbital properties, the subsequent pairing of two collision products inevitably requires dynamical processes, such as three-body and binary--single interactions. Hence, it is reasonable to assume that BBHs with both components in the upper-mass gap formed in a dynamically active environments, i.e. a star cluster or galactic nucleus. The spins of BHs formed through stellar mergers are hard to constrain, owing to the large uncertainties on both the natal spins distribution and the impact of stellar structure on the spins of the remnant BHs. 
Hierarchical BBH mergers, instead, can occur only if  merger remnants are retained in the host cluster, i.e. if the host escape velocity is larger than the relativistic kick imparted to merger remnants \citep{campanelli_2007, gonzalez_2007, lousto_2008, lousto_2012}. 
In such a case, remnants can further pair-up and merge with other BHs, or among themselves, and build-up heavier BHs \citep[]{Antonini_2019, Mapelli2021,zevinOneChannelRule2021a,Arca_Sedda_2023, Kritos2024, Torniamenti2024, alvarez_2024, mahapatra_2024}. This channel may be particularly efficient in active galactic nuclei, where the gravitational potential of the central supermassive BH keeps BBH merger remnants bound to the galactic center \citep{2012MNRAS.425..460M,2020ApJ...898...25T,2023Univ....9..138A}.
The latter have final spins $\sim 0.7$ \citep{Berti2007, Hofmann2016}, although the spin amplitude tends to decrease in the case of long merger chains \citep{Miller2002,2003ApJ...585L.101H}. 

In this letter, we use the \bpop semi-analytic population synthesis code to explore the formation of BBH mergers similar to GW231123 forming via dynamical interactions in young (\young), globular (\globular), and nuclear star clusters (\nuclear), keeping into account both the stellar mergers and hierarchical BBH merger channels. 

The letter is organized as follows. In Section \ref{sec:methods} we discuss our methodology. In Section \ref{subsec:massdist} we investigate the BBHs mass distribution in various environments and the fraction of mergers compatible with GW231123. In Section \ref{subsec:origins_spins}, we study the effect of different natal spin prescriptions for BHs in the upper-mass gap, and in Section \ref{subsec:delay_times} we infer the formation redshift of GW231123-like progenitors, discussing potential host environments and related merger rates. Finally,  in Section \ref{sec:conclusions} we summarize our results.

\setcounter{footnote}{0}

\section{Methods}
\label{sec:methods}
The astrophysical population of BBH mergers is generated utilizing the \bpop semi-analytic population synthesis tool \citep[]{ArcaSedda_Benacquista_2019,arca_sedda_fingerprints_2020,Arca_Sedda_2023}, which allows to simulate an heterogeneous populations of BBH mergers originating from different astrophysical environments, namely isolated fields and YCs, GCs, and NCs. \bpop samples BH natal masses from pre-compiled catalogs generated with dedicated stellar evolution codes. In this work, we utilise a catalog generated with the state-of-the-art binary population synthesis code \textsc{SEVN} \citep{Spera2015,speraMergingBlackHole2019a, iorioCompactObjectMergers2023}, following the fiducial assumptions detailed in \cite{iorioCompactObjectMergers2023}. \bpop combines the catalogs to sample the masses of BHs either formed from the collapse of single stars or in a binary system. The latter scenario permits us to
take into account the effects of {\it primordial} stellar binaries, which can constitute a significant fraction of observed young stellar populations \citep{Moe_2017} and can significantly shape the BH mass spectrum \citep{speraMergingBlackHole2019a}. In particular, stellar mergers and collisions\footnote{Throughout this work, we refer to any process that increases a BH’s natal mass in a stellar binary beyond the upper-mass gap as a “stellar merger/collision”, both for conciseness and because most BHs in the gap form this way.} as well as mass transfer in close binaries, can significantly affect the BH mass spectrum, populating the upper-mass gap \citep{speraMergingBlackHole2019a, Arca_Sedda_2023}.
In \bpop, BBH dynamics is regulated by three-body and binary-single scatterings, which contribute to harden the binary up to the point that it either merges inside the cluster or is ejected away. When a BBH merges inside the cluster, its remnant can be recoiled due to its relativistic kick, unless the kick is lower than the cluster's escape velocity. 
Retained merger remnants can undergo additional mergers, leading to hierarchical BBH merger chains.
Note that, in its current version, \bpop follows only primary BHs along the chain, assuming that any secondary BH is a first generation BH either formed from a single star or in a binary. We refer the reader to \cite{Arca_Sedda_2023} for a detailed description of the methodology implemented in the code to model BBH dynamics.
In the fiducial model, natal spins are drawn from a Maxwellian distribution with dispersion $\sigma_\chi = 0.2$, as suggested by LVK observations performed during the first three observing runs \citep{theligoscientificcollaborationPopulationMergingCompact2022}. While this constrain is also supported by theoretical models of BH formation from single stars \citep{Fuller2019}, BHs forming in binary systems can be spin-up through episodes of matter accretion or tidal spin-up \citep[e.g.][]{Qin2018}, and those forming from stellar mergers may retain angular momentum that is not efficiently dissipated during the collision and collapse of the product \citep{Ryu2023}. 
Despite the post-merger recoil depends in a non trivial way on the spin amplitude and orientation, in general the larger the spins the higher the maximum kick the remnant can receive \citep[see e.g. Figure 3 in][]{Antonini_2016}. Hence, 
large natal spins can significantly impact the properties of BHs formed through hierarchical mergers, possibly offering a powerful tool to interpret GW observations, despite the low accuracy associated with spin measurements.

Each BBH is co-evolved with its own cluster, which contracts and expands due to relaxation. Cluster dynamical evolution is modelled through fitting formulae that reproduce the dynamical evolution of the \dragon\, simulations database, a suite of 19 state-of-the-art direct \textit{N}-body models of massive YCs (see \citealt{AS_DragonII_1, AS_DragonII_2, AS_DragonII_3}). This enables the coupling of the complex dynamics of both BBHs and their host star clusters.

In this letter, we use \bpop to simulate BBH mergers in YCs, GCs, and NCs, assuming three values of the stellar metallicity, namely $Z = 0.0002,~0.002,~0.02$, and that a fraction 
$f_{\rm bin} = 0.6$ of BHs participating in merging events formed in stellar binaries. This choice reflects the fact that most BH progenitors, i.e., massive stars, are born in binary systems \citep[see e.g.][]{Moe_2017}. 
For each combination, we simulate $5 \times 10^7$ BBHs, i.e. in total $N_{\rm sim} = 4.5 \times 10^8$ BBHs. Note that only a fraction $f_{\rm sim}$ of all BBHs will merge within a Hubble time, depending on binary orbital properties, stellar metallicity, and star cluster structure. In particular, we find that the fraction of BBHs that merge within a Hubble time is $f_{\rm sim} \sim 0.02-0.07$ for \young, $\sim 0.14 - 0.40$ for \globular, and $\sim 0.35 - 0.74$ for \nuclear, with the largest (smaller) fractions corresponding to lower (higher) metallicities.

\section{Results}
\label{sec:results}
The main outcomes of the simulated population of BBH mergers with properties similar to GW231123 are discussed in the following. 
Throughout the section, we classify BBH mergers in three categories: (i) BSE (binary stellar evolution), if their components never underwent previous BBH mergers and at least one formed in a stellar binary with a mass in the upper-mass gap; (ii) H (hybrid), if the primary BH undergoes at least one previous merger and, like in the BSE case, either its initial mass falls in the upper-mass gap and/or one secondary BH along the merger chain formed in the gap
; (iii) MCs (merger chains), if their primary had an initial mass below the upper-mass gap, and it solely grew via hierarchical mergers with BHs below the gap.
Hence, to summarize, BSE-BBH mergers are not part of hierarchical chains but involve upper-mass gap BHs, MCs-BBH mergers are hierarchical but do not involve any BHs born in the gap from primordial stellar binaries, while H-BBH mergers combine both hierarchical BBH mergers and BHs produced in the gap.
\subsection{Massive BBHs in star clusters: the combined roles of stellar mergers and hierarchical BBH mergers}
\label{subsec:massdist}

\begin{figure*}[htp]
    \centering
    \includegraphics[width=1.8\columnwidth]{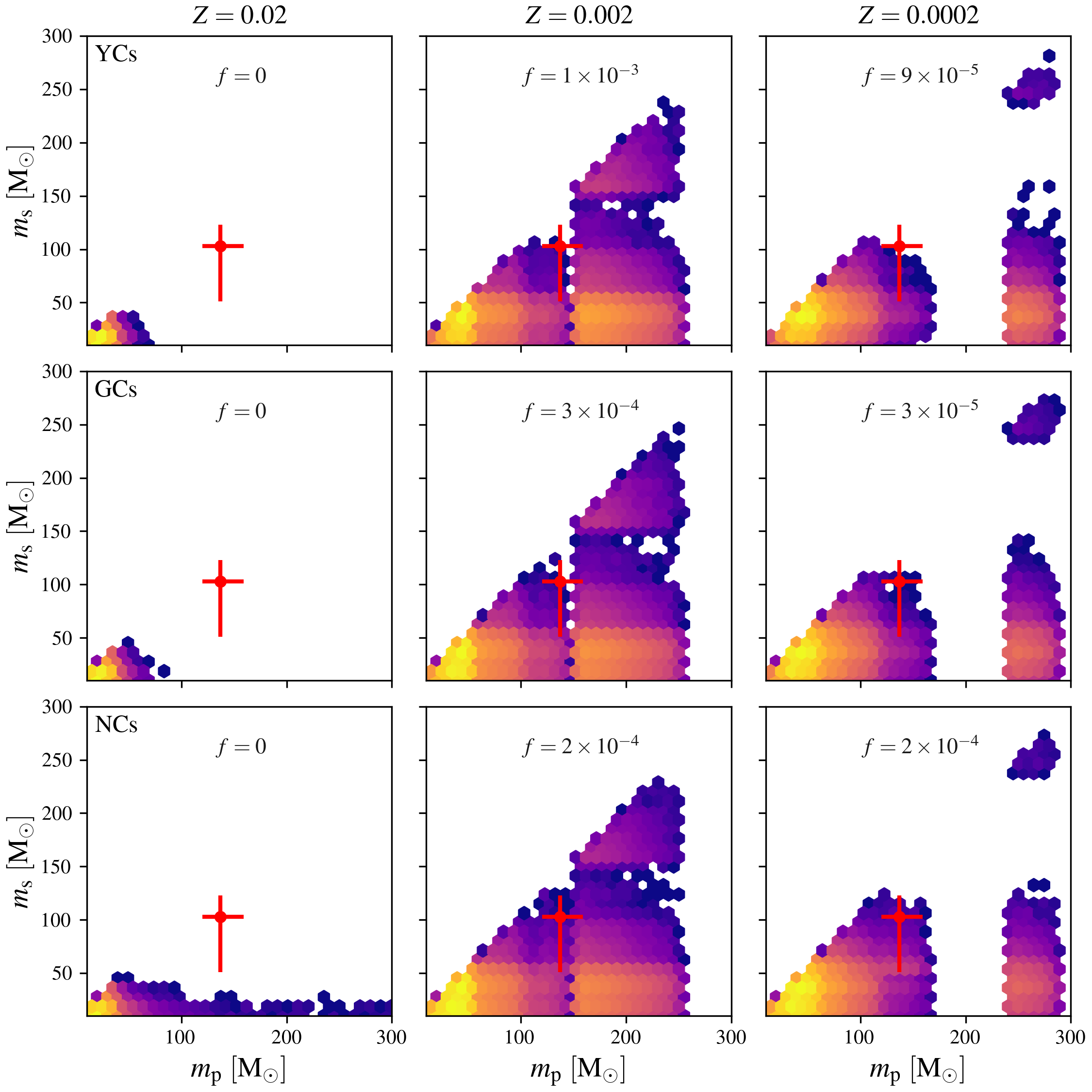}    \caption{BBH populations for different metallicities($Z=0.02$ on the left column, $Z=0.002$ on the central column, and $Z=0.0002$ on the right column) and cluster environments, namely \young (upper row), \globular (central row), and \nuclear (bottom row). The bins are normalized over the number of BBHs with brighter (darker) colors corresponding to more (less) BBHs. GW231123 primary and secondary masses are indicated with a red cross. 
    The fraction of BBHs falling within the event mass ranges is indicated for each metallicity–environment combination in the plots.
    }
    \label{fig::fiducial_grid}
\end{figure*}
Figure \ref{fig::fiducial_grid} shows the combined primary mass - secondary mass distribution for the set of fiducial models with different metallicities and environments. For each combination we present the fraction of GW231123-like BBH mergers calculated as the ones falling within the 90 \% credible interval (C.I.) of GW231123 primary and secondary masses \citep{GW231123_paper}.
\par
Comparing the various panels highlights the crucial impact of metallicity in producing GW231123-like binaries. Models with $Z = 0.02$ never form a source with component masses similar to GW231123, while more metal-poor environments have the chance to form such type of sources in all the types of clusters explored. 
Table \ref{tab:bbh_origins} summarizes the main results of this comparison.  We find a fraction of compatible merger events $f_{\rm mer} = (0.3-10)\times 10^{-4}$. 
Regardless of the cluster type, the highest fraction of mergers compatible with GW231123 masses is reached for a relatively small metallicity value, namely $Z = 0.002$. In general, we expect an increase of BBH mergers in GW231123 mass range at low metallicities. 
 Indeed, metal-poor stars lose less mass during their lifetime, leading to more massive BHs \citep{belczynski_comprehensive_2002, Heger2002,fryer_compact_2012, sukhbold_core-collapse_2016,speraVeryMassiveStars2017, limongi_presupernova_2018, vinkPresupernovaEvolutionExplosive2018,costaFormationGW190521Stellar2021, iorioCompactObjectMergers2023}. 
Additionally, the inclusion, in our runs, of stellar merger products formed from binary star evolution further facilitates the formation of BHs in and above the upper-mass gap. This implies that the mass spectrum of BHs in the gap formed from binary stars crucially depends on both (P)PISN physics \citep{woosleyPulsationalPairinstabilitySupernovae2017, Farmer2019, renzoPredictionsHydrogenfreeEjecta2020, hendriksPulsationalPairinstabilitySupernovae2023, Ugolini2025} and binary properties \citep{marchantPulsationalPairinstabilitySupernovae2019a, costaMassiveBinaryBlack2023}, two ingredients that can significantly affect the formation of massive BBH mergers.

In Appendix \ref{app:gapdependence} we briefly examine the connection between the BH mass spectrum produced by stellar binaries and the resulting fractions of GW231123-like systems in our simulations. 

The relatively high value of $f_{\rm mer}$ in YCs owes to the fact that these are relatively short-lived systems (with lifetimes $\sim 1-5$ Gyr) that promote the formation of more massive BBHs, since they have shorter evolution timescales. GCs and NCs, with lifetimes $\gtrsim 10$ Gyr, can form also much lighter BBH mergers, thereby reducing the fractional number of possible GW231123-like mergers. 

\begin{figure}[h]
    \centering
    \includegraphics[width=\columnwidth]{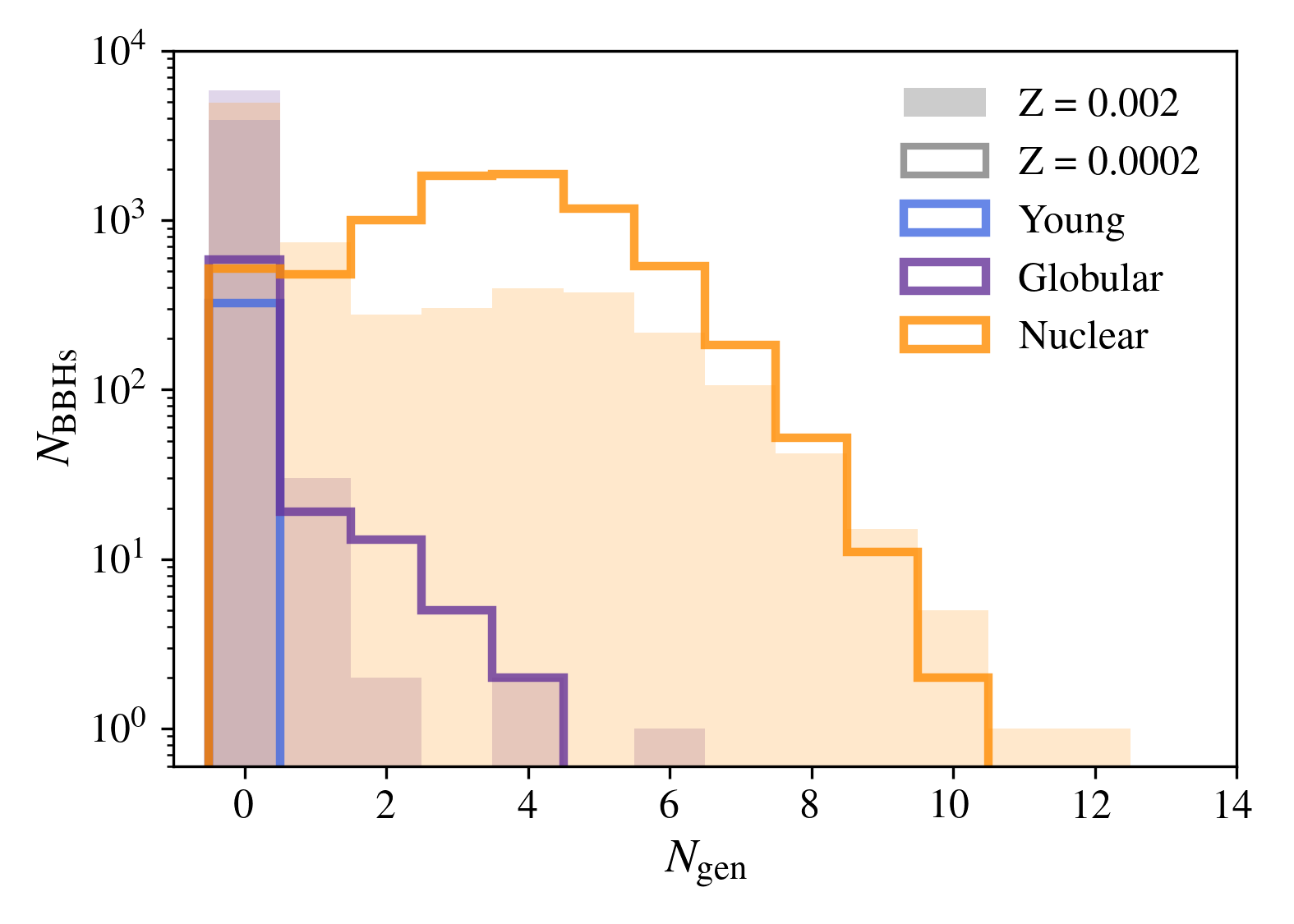}    
    \caption{Distribution of primary BHs generations for simulated BBH mergers with component masses in 90 \% C.I. of GW231123. We distinguish between models with a metallicity $Z = 0.002$ (filled steps) or $Z=0.0002$ (open steps), and among \young (blue steps), \globular (purple steps), and \nuclear  (orange steps). }
    \label{fig::fiducial_grid_Ngen}
\end{figure}

Note that, according to the stellar evolution setup adopted in our work, 
GW231123's BHs are too heavy to be produced solely via single stellar evolution (see more in Appendix \ref{app:gapdependence}).
This implies that only simulated BHs formed through binary stellar processes or repeated BBH mergers can reach such masses. 
To discern between massive BBHs from single mergers and from hierarchical chains, we show in Figure \ref{fig::fiducial_grid_Ngen} the distribution of merger generations for simulated BBH mergers with component masses compatible with  GW231123. Note that the generation of a BH represents the number, $N_{\rm gen}$, of previous mergers it underwent. Clearly, BHs that never merged before, and likely formed from stellar mergers in primordial binaries, are characterized by $N_{\rm gen} = 0$, whereas BHs that are byproduct of hierarchical mergers are characterized by $N_{\rm gen} > 0$.
Hence, Figure \ref{fig::fiducial_grid_Ngen} highlights the crucial role of stellar mergers in determining the formation of GW231123-like systems. In fact, we see that $N_{\rm gen} \sim 1-6$ only for a handful of BBH mergers from GCs, and $N_{\rm gen} = 0$ for all mergers from YCs. 
This owes to the large relativistic kicks imparted to merger products, which can be as large as $v_{\rm GW} \sim 10^3-10^4$ km$/$s, i.e. much larger than the typical escape velocity of YCs, $v_{\rm esc} \sim 1-5$ km$/$s, and GCs, $v_{\rm esc} < 50$ km$/$s \citep[see e.g. Figure 8 in][]{arcasedda_2021}. Additionally, in YCs and GCs the ejection of BBHs can also occur through strong gravitational interactions, which can impart Newtonian recoils up to $\sim 10-100$ km$/$s \citep[e.g.][]{Sigurdsson1995, AS_DragonII_2}.

In NCs, which can have escape velocities up to several $10^2$ km$/$s (see Figure 1 in \cite{Antonini_2016} and Figure 8 in \cite{arcasedda_2021}), 
BSE- and H-BBH mergers are counterbalanced by MC-BBH mergers. As shown in Figure \ref{fig::fiducial_grid_Ngen}, in these clusters the fraction of BBHs from hierarchical chains that reach the GW231123 mass range increases as metallicity decreases, with some BHs undergoing up to $\sim 10–12$ mergers.

\subsection{The fundamental impact of spins}
\label{subsec:origins_spins}

\begin{table*}[t]
  \footnotesize
  \caption{\small BBH mergers per cluster type and metallicity. Col. 1: Cluster type. Col. 2: Metallicity. Col 3: fraction of mergers compatible with GW231123's masses in the low-spin model. Col. 4: fraction of mergers compatible with GW231123's masses and spins in the low-spin model. Col. 5-8: number of BSE-BBH mergers in the low-spin model. Col. 9-11: same as before, but for H-BBH mergers. Col. 12-14: same as before, but MC-BBH mergers.}
\label{tab:bbh_origins}
  \begin{tabular*}{\textwidth}{@{\extracolsep{\fill}}
      l c c c
      *{3}{c}
      *{3}{c}
      *{3}{c}
    @{}}
    \toprule
    Clu. & Z 
         & \multicolumn{1}{c}{$f_{\mathrm{mer}}$} 
         & \multicolumn{1}{c}{$f_{\mathrm{mer},\chi}$} 
         & \multicolumn{3}{c}{BSE} 
         & \multicolumn{3}{c}{H} 
         & \multicolumn{3}{c}{MC} \\
    \cmidrule(lr){3-3} \cmidrule(lr){4-4} \cmidrule(lr){5-7} \cmidrule(lr){8-10} \cmidrule(lr){11-13}
    & & low & low 
      & low & mid & high
      & low & mid & high
      & low & mid & high \\
    \midrule
    \multirow{2}{*}{YCs}
      & 0.002   & $1\times10^{-3}$ & $3\times10^{-6}$ & 10 & 435 & 2250 &  -- &  -- &  -- &  -- &  -- & -- \\
      & 0.0002  & $9\times10^{-5}$ & $3\times10^{-7}$ &  1 &  24 &  145 &  -- &  -- &  -- &  -- &  -- & -- \\
    \midrule
    \multirow{2}{*}{GCs}
      & 0.002   & $3\times10^{-4}$ & $2\times10^{-6}$ & 21 & 572 & 3368 &   1 &   -- &   -- &   4 &   4 &  4 \\
      & 0.0002  & $3\times10^{-5}$ & $3\times10^{-7}$ &  2 &  54 &  274 &  -- &  -- &  --  &   4 &   4 &  4 \\
    \midrule
    \multirow{2}{*}{NCs}
      & 0.002   & $2\times10^{-4}$ & $8\times10^{-6}$ & 21 & 478 & 2787 &  57 &  30 &   28 & 182 & 182 & 182 \\
      & 0.0002  & $2\times10^{-4}$ & $2\times10^{-4}$ &  3 &  48 &  208 &  37 &  16 &   15 & 845 & 845 & 845 \\
    \bottomrule
  \end{tabular*}
\end{table*}

In the previous section, we restricted the analysis to BBH mergers with component masses compatible with GW231123 C.I.. In this section, we add a further element of comparison, namely the spins of the two BHs, $\chi_{1,2}$, and the effective spin parameter \citep[][]{Arun2009, Reisswig2009, Ajith2011}, given by the projections of the two spin vectors onto the binary angular momentum vector, i.e.
\begin{equation}
    \chi_{\rm eff} = \displaystyle\frac{\left(m_1\vec{\chi_1}+m_2\vec{\chi_2}\right)\cdot\vec{L}}{m_1+m_2}.
\end{equation}

The inferred spin values for GW231123 suggest that both BHs are highly spinning, exhibiting $\chi_1 = 0.90^{+0.10}_{-0.19}$ and $\chi_2 = 0.80^{+0.20}_{-0.51}$. 
High spins are generally considered a signature of multiple mergers, especially of first generations mergers involving comparable mass BHs. In such cases, the remnant spins is typically $\chi_{\rm rem} \sim 0.7$ \citep{Berti2007, Hofmann2016}, a value largely determined by General Relativity effects. Variations in the mass ratio and in the magnitudes and orientations of the progenitors spins can produce scatter around this characteristic value (see e.g. Figure 1 in \cite{alvarez_2024}).
The fractions of mergers with both component masses and spins falling inside the $90\%$ C.I. for the event, $f_{\rm mer, \chi}$, are listed in Table \ref{tab:bbh_origins} for different models.

The additional condition on the BHs spins decreases the fraction of BBHs able to produce GW231123-like events typically by one or two orders of magnitudes.
The small fraction of BBH mergers falling in the GW231123 mass-spin quadridimensional box is mostly due to the adopted prescription for natal spins of BHs formed from stellar processes in binary systems, i.e. the Maxwellian with $\sigma_{\chi} = 0.2$, which we refer to as the ``low-spin'' model. 
However, BHs forming above the upper-mass gap due to accretion episodes in binary systems or stellar mergers can also gain high spins, up to extremal values, especially if they are the second-born BH in a binary (see \citealt{Qin2018}). Unfortunately, natal BH spins are still poorly constrained by stellar evolution theories, especially in the case of those forming from stellar mergers, thus a solid theoretical framework to model this quantity is still missing. To assess the impact of different spin prescriptions on our results, we resample the spins of BHs in the gap of BSE-BBHs
from a Gaussian distribution centered either on $\chi_{\rm med} = 0.5$ (hereafter ``mid-spin'' model) or $\chi_{\rm med}= 0.9$ (hereafter ``high-spin'' model), assuming a dispersion of $\sigma_{\rm \chi} = 0.2$. Note that this sampling procedure does not affect the merger probability of BHs forming through stellar mergers, collisions, and accretion in binaries. 
In the case of H-BBH mergers, instead, resampling the spin of one event in the chain can have a catastrophic effect on all the following mergers. Larger spins may lead to stronger kicks, leading to premature ejection of the BH remnant from the host cluster or to a lengthening of the dynamical timescales needed for the BH remnant to pair with another BH.
Therefore, in this case we resample only the spin of the first BH along the chain of mergers which is born in the upper-mass gap.
We then assess the retention of the BBH merger product and calculate the number of GW231123-like BBHs in the new sample, to be compared with the number of mergers found in the low-spin model. 
In Table \ref{tab:bbh_origins}, the number of BBHs falling within the BSE, H, or MC category are summarized for the low-, mid-, and high-spin models. 
Increasing the spin of BSE-BHs boosts by two (three) order(s) of magnitude the number of mergers compatible with GW231123 in the mid-(high-)spin model, relative to the low-spin model, regardless of the environment. This enhances the probability for \globular and \young to nurture the formation of GW231123-like systems. 

Moreover, increasing the spins reduces the probability for hierarchical mergers involving upper-mass gap BHs to occur. 
Indeed, the number of H mergers compatible with GW231123 reduces by  $\gtrsim 50\%$ in \nuclear. MC-mergers are not influenced by the change in spins adopted since, by definition, they never involve BHs formed in or above the gap. 
Note that the spin re-sampling approach adopted for hierarchical mergers may be conservative, but enables us to highlight the fundamental role of spin amplitudes in determining the successful development of a merger chain. 

Our models clearly highlight a key difference between BBH mergers forming in different environments. In YCs and GCs, the efficiency of hierarchical mergers is so poor that GW231123-like systems can form practically only through the BSE channel. 
Conversely, in NCs, H- and MC-mergers can account for almost all GW231123-like mergers in the low spins scenario, though the relative contribution of this channel can vary significantly depending on the natal spin distribution of stellar merger products and on the metallicity.

\begin{figure}[h]
    \centering
    \includegraphics[width=\columnwidth]{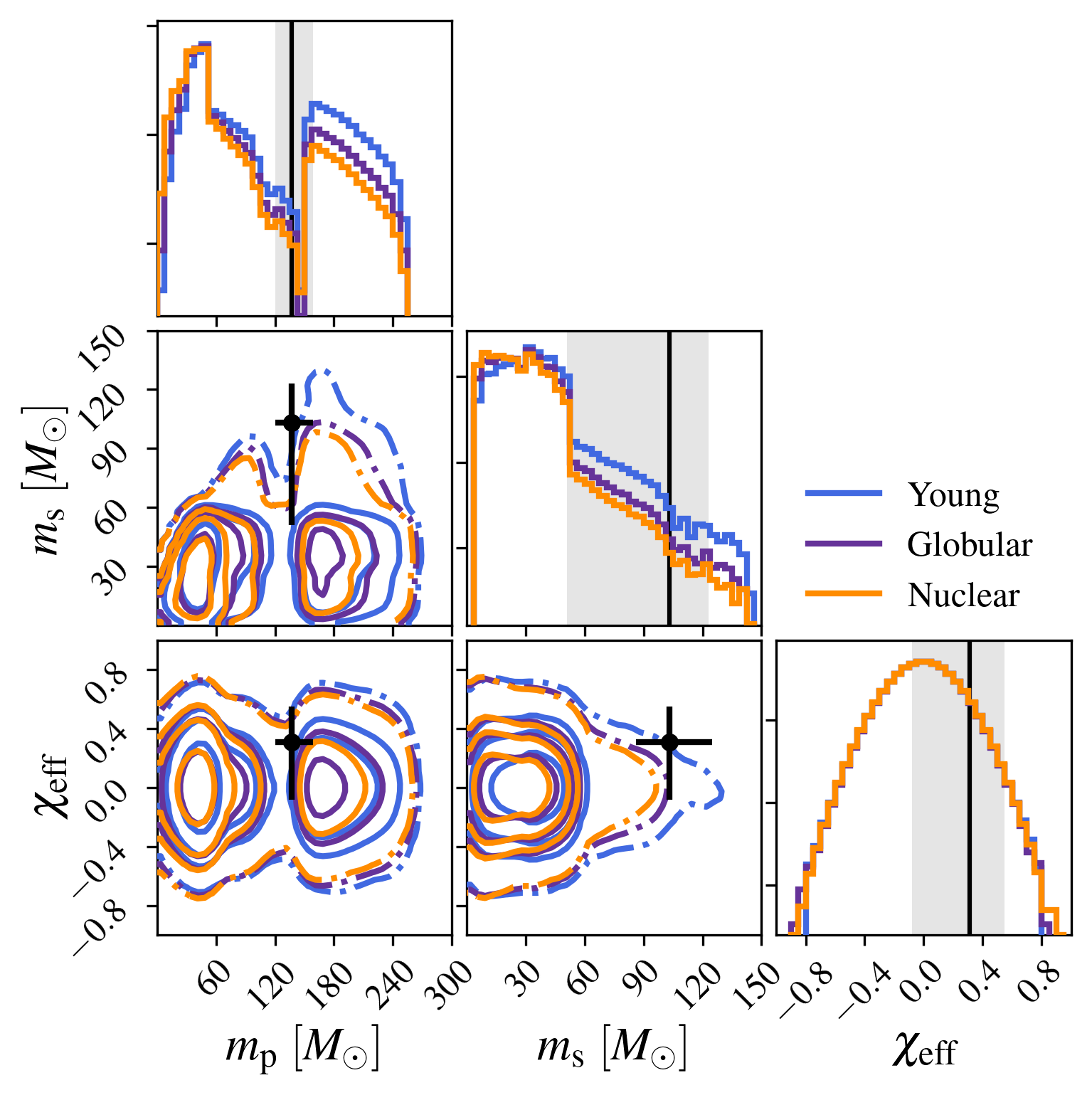}
    \caption{Masses and effective spin distributions for \young, \globular and \nuclear in our fiducial simulation. The metallicity is set to $Z = 0.002$. The contour lines refer to the 68\%, 95\%, 99 \% (solid lines) and 99.99 \% (dotted-dashed line) contours. The 1-D distributions are normalized to 1 and displayed in log-scale.}
    \label{fig3:corner}
\end{figure}

\begin{figure}[h]
    \centering
    \includegraphics[width=\columnwidth]{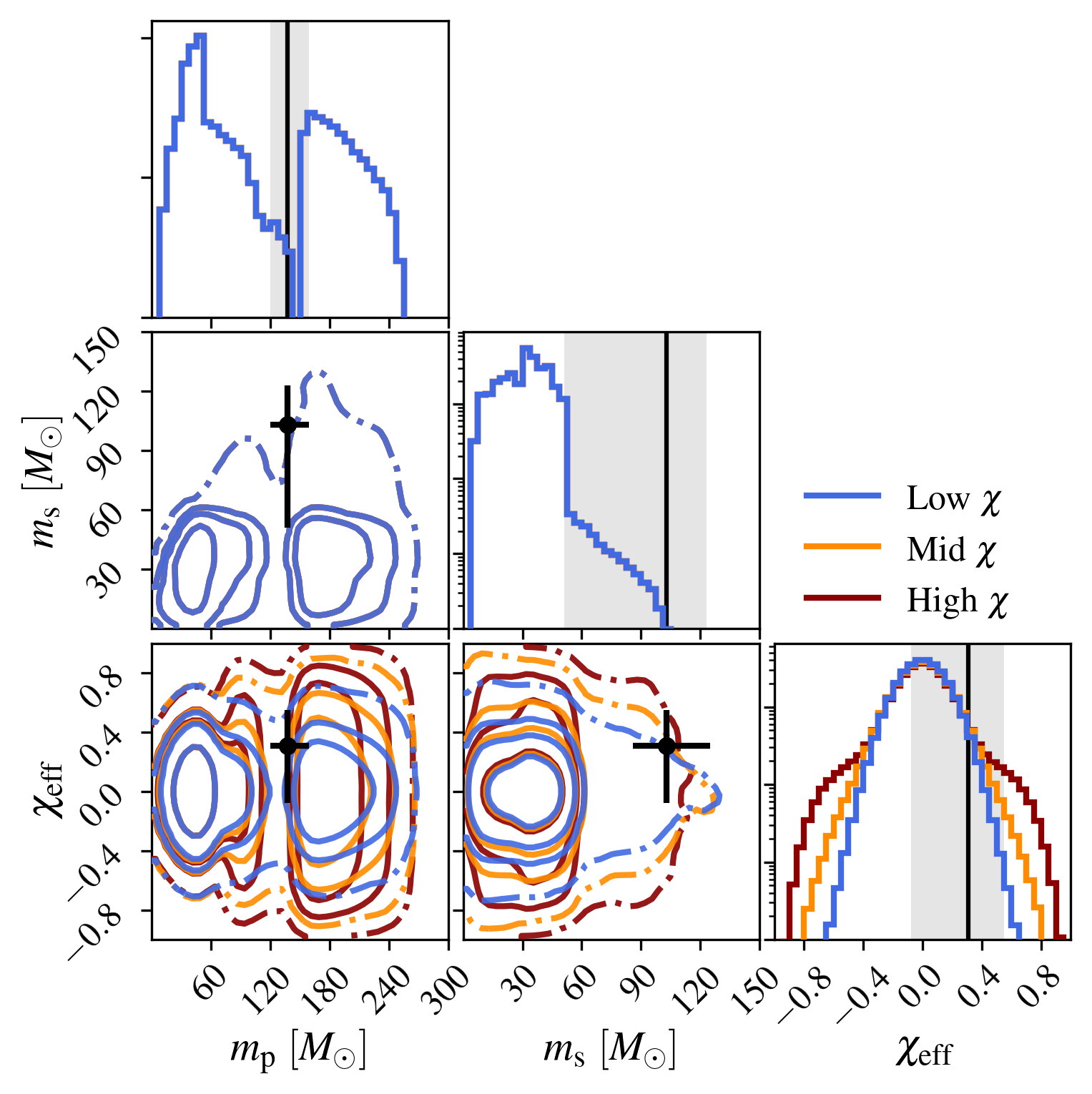}
    \caption{Masses and effective spin distributions for stellar BBHs in \young in our fiducial simulation. The metallicity is set to $Z = 0.002$.} The contour lines refer to the 68\%, 95\%, 99 \% (solid lines) and 99.99 \% (dotted-dashed line) contours. The 1-D distributions are normalized to 1 and displayed in log-scale.
    \label{fig4:corner}
\end{figure}

We summarize the main properties of simulated mergers in a corner plot displaying masses and effective spin parameters for models with $Z=0.002$ and different environments (Figure \ref{fig3:corner}), or different spins prescriptions and \young only (Figure \ref{fig4:corner}). The two plots highlight several key findings, which can be summarized as follows:
\begin{itemize}
    \item our models are able to produce the primary mass of GW231123 in all types of environment (first column of Figure \ref{fig3:corner}), especially if the primary BH formed from stellar mergers and had high spin (bottom left panel Figure \ref{fig4:corner}), although BBHs with such a primary falls in the high-end tail of the mass distribution;
    \item our models cannot fully capture the properties of the companion, specifically the high-end tail of the mass distribution (second row of Figure \ref{fig3:corner}, and Figure \ref{fig4:corner}). This can owe to the adopted stellar evolution paradigm, or the assumption that all secondary in BBH mergers in \bpop did not undergo any previous merger, an assumption well supported by numerical simulations \citep[see e.g.][]{Rodriguez2019,AS_DragonII_2} that however may fail in the densest clusters \citep{Kritos2024};
    \item within our fiducial model, our analysis suggest that BHs forming from the collisions of stars in binaries should have significant natal spins to accommodate the inferred spin distribution of GW231123 components (last row of Figure \ref{fig4:corner}).  
\end{itemize}

It is worth mentioning that the inference of GW231123 parameters assumes the binary orbit to be quasi-circular. However, accounting for eccentric waveforms could significantly impact the spin parameter estimates and potentially provide a better fit to the data, as discussed for the GW190521 event in several previous studies \citep[see e.g.][]{ecc_gayathri_2020, ecc_romeroshaw_2020, ecc_bustillo_2021}.

\subsection{Delay times distribution and potential host galaxies}
\label{subsec:delay_times}
The redshift at which GW231123 has been observed, $z_{\rm mer}= 0.39^{+0.27}_{-0.24}$, can also represent a valuable source of information on the event's host formation history. 
This redshift value corresponds to an age of the Universe\footnote{assuming Planck18 cosmology \citep{2020A&A...641A...1P}.} of $t_{\rm age,  mer} \sim 9.45 $ Gyr. 

\begin{figure}[h]
    \centering    \includegraphics[width=\columnwidth]{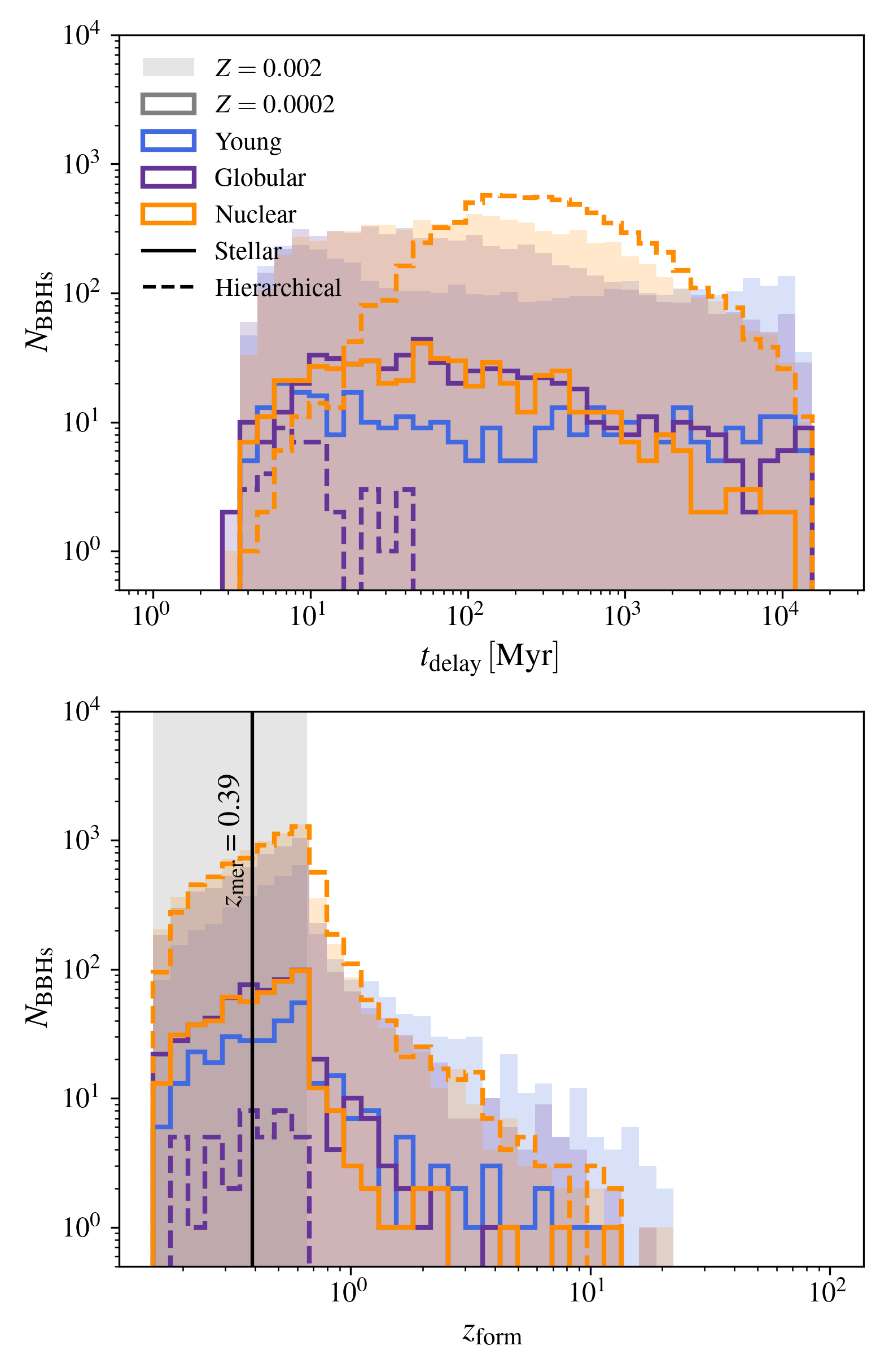}
    \caption{Delay times of mergers in the GW231123 masses C.I. for different cluster environments in our simulations. The filled histograms refer to $Z = 0.002$ runs, while the empty histograms refers to $Z = 0.0002$ runs. In the latter case, the solid lines refer to primary BHs undergoing a single merger event (Stellar) while the dashed lines refer to BHs which underwent also previous mergers (Hierarchical). 
    }
    \label{fig:delaytimes}
\end{figure}

In Figure \ref{fig:delaytimes} upper panel, we present the delay times, $t_{\rm delay}$, of all GW231123-like mergers, i.e. the total time elapsed from the formation of the two BHs, to their pairing, hardening, and merger. The delay time distributions are shown for different environments and metallicities. For $Z = 0.0002$, we also highlight the separate contributions of BSE-BBHs (solid line) and H- and MC-BBHs (dashed line). In the case of BSE-BBHs, the distribution is nearly flat in logarithmic values between $t_{\rm delay} = 10^6-10^{10}$ yr, with the vast majority of mergers occurring within $t_{\rm delay} < 1 $ Gyr, regardless of the environment or metallicity. The distribution for H- and MC-mergers, instead, exhibits a clear peak around $0.1-1$ Gyr driven by the multiple dynamical processes concurring to the build-up of merger chains. 

For each BBH merger in our simulated catalogs, we calculate the age of the Universe at which the binary must form to merge at the observed redshift, i.e. $t_{\rm form} = t_{\rm age, mer}(z_{\rm mer})$, sampling the merger redshift $z$ randomly in the 90 \% C.I. reported for the event \citep{GW231123_paper}. We convert the age into a formation redshift $z_{\rm form} = z\,(t_{\rm age, mer} - t_{\rm delay})$, which represents the redshift at which the stellar progenitors formed in the cluster. The bottom panel of Figure \ref{fig:delaytimes} compares formation redshifts and the GW231123 merger redshift. 

Given the short delay time, our models suggest that the formation redshift of GW231123-like systems does not differ much from the redshift at merger, although all models exhibit a significant tail extending out to redshift $z_{\rm form}>6$.
This type of approach can help placing constraints on the properties of the environment at the time of the BBH merger formation, and possibly to the cluster formation history. 
For example, there is a clear correlation between galaxies' stellar mass and metallicity, the so-called fundamental mass-metallicity relation \citep[]{Mannucci2010, Curti2020}, which highlights how, on average, the more massive the galaxy, the higher the metallicity. In the local Universe, both observations and numerical models suggest that metallicity below $Z < 0.002$ can be found only in dwarf galaxies, with stellar masses $\sim (10^6-10^8)\msun$ \citep{2013ApJ...779..102K,2016MNRAS.456.2140M, 2017A&A...601A..95C}. Alternatively, GCs offer a naturally metal-poor environment for the formation of GW231123-like sources. The Galactic population of GCs, for example, features a bi-modal metallicity with the majority of them having $Z \lesssim 0.1$ Z$_\odot$ \citep[e.g.][]{1984ApJS...55...45Z}, a feature recovered also in other extragalactic GC populations \citep{2006ARA&A..44..193B}. Also Galactic \nuclear exhibit a metal-poor population accounting for about $7-10\%$ of all the stars in the Galactic Center \citep{2020ApJ...901L..28D}, possibly reminder of a massive cluster spiraled there a few Gyr ago \citep{2020ApJ...901L..29A}. 

Given the generally low efficiency of hierarchical mergers and the strong impact of (P)PISN in environments with solar metallicity, our models suggest that the most likely birth site of GW231123 is in a metal-poor cluster orbiting a dwarf galaxy located at a redshift $z < 1$. 

While we postpone a thorough discussion about the merger rate evolution of this type of mergers, in the following we briefly outline an order of magnitude calculation to place constrain on the possible origin of this source. 

The local number density of galaxies similar to the Milky Way is $n_{\rm glx} \simeq 0.0116$ Mpc$^{-3}$ \citep{2008ApJ...675.1459K}. The corresponding value of $n_{\rm dwr} \sim 10 \,n_{\rm glx}$ should be an order of magnitude larger, according to galaxy formation theories. In a typical dwarf galaxy, we expect at most $N_{\rm clu} = 1$ \nuclear, and $N_{\rm clu} = 10$ \globular and \young. In a Milky Way-like galaxy, instead, we observe around $N_{\rm clu} = 200$ \globular \citep{1996AJ....112.1487H,2010arXiv1012.3224H,2017ApJ...849L..24M} and $N_{\rm clu} \sim 10^3$ massive \young \citep{larsen_2009}, although most of the latter exhibit solar or super-solar metallicities \citep{2013A&A...558A..53K}. 
The total number of BHs in a stellar population is given by $n_{\rm bhs} \sim 10^{-3} N_{*, \rm clu}$, with $N_{*, \rm clu}$ representing the total number of elements in the cluster stellar population. We dub with $\epsilon_{\rm mer}$ the fraction of BHs which end up producing a merger over the number of BHs simulated. We evaluate this efficiency term directly from our simulations and get that it is 0.06-0.07 for \young, 0.35-0.40 for \globular and 0.65-0.74 for \nuclear for $Z = 0.002$ and $Z = 0.0002$ respectively. Hence, the effective number of BBH mergers per cluster would be given by $N_{\rm BBH, mer} = \epsilon_{\rm mer} \ n_{\rm bhs} \ N_* $. To get the effective number of mergers compatible with GW231123's masses per cluster we need to multiply $N_{\rm BBH, mer}$ by the $f_{\rm mer}$ in Table \ref{tab:bbh_origins}. For both $\epsilon_{\rm mer}$ and $f_{\rm mer}$ we consider the mean value across the two metallicity considered. Finally, we assume that the delay times of the BBH mergers lie in the range $t_{\rm del} \sim 0.1$--$1 \, \mathrm{Gyr}$ (see Figure \ref{fig:delaytimes} top panel).

Hence, we can roughly estimate the rate of GW231123-like events from \young, \globular and \nuclear as
\begin{equation}
        \mathcal{R} = \frac{f_{\rm mer} \ \epsilon_{\rm mer} \ n_{\rm bhs} \ N_{\rm clu}\ n_{\rm dwr}}{t_{\rm del}} \ .
\end{equation}
In the case of dwarf galaxies, this approach leads to a local merger rate of
\begin{equation}
    \mathcal{R}_{\rm DW}\sim 10^{-2}  {\rm yr^{-1}~Gpc}^{-3} \times
    \begin{cases}
         (1.6-16) & {\rm NC},\\
         (3.6-36)  & {\rm GC},\\
         (0.04-0.41) & {\rm YC} .    
    \end{cases}
    \label{eq3:rate}
\end{equation}
where we simply assume $N_* = 10^6-5\times10^5-10^4$ for \nuclear, \globular, and \young, respectively. 
Similarly, for Milky Way-like galaxies we can make the same calculation using $n_{\rm glx}$ to obtain
\begin{equation}
    \mathcal{R}_{\rm MW}\sim  10^{-2}  {\rm yr^{-1}~Gpc}^{-3} \times
    \begin{cases}
         (0.16-1.6) & {\rm NC},\\
         (7 - 72) & {\rm GC},\\
         (0.4-4.1) & {\rm YC}.
    \end{cases}
    \label{eq4:rate}
\end{equation}

The estimated rate in Eqns. \eqref{eq3:rate} and \eqref{eq4:rate} rely on the assumption that all environments considered are metal-poor. This requirement is generally satisfied in GCs \citep{1996AJ....112.1487H,2010arXiv1012.3224H}, and NCs, which host sub-population of old, metal-poor, stars \citep{2020A&ARv..28....4N,2020ApJ...901L..28D}. The metal content of YCs is less constrained, but the limited available observations suggest they preferentially host metal-rich stellar populations \citep{Portegies_Zwart_2010}. Therefore, for YCs our estimate may represent an upper limit to the real rate.
Nonetheless, several uncertainties can affect the estimated rate, like the fraction of metal-poor dwarf galaxies, the dependence between the galaxy mass and the number of GCs and YCs forming there, the metallicity spread within single galaxies, or the initial mass function (IMF) of the stellar population.
While the general consensous suggests that stellar population may distributed according to a Universal IMF \citep{Kroupa2001}, in the last few years several works suggested that in metal-poor environments stellar formation can sustain a top-heavy IMF \citep[e.g.][]{marks_2012}. 
At the lowest level of approximation, a top-heavy IMF can increase the fraction of binary stars that lead to the formation of GW231123 components, readily translating into a proportional increase in the estimated merger rate.

The impact of a top-heavy IMF on BHs undergoing hierarchical mergers is somewhat more difficult to predict. Changes in the BH mass spectrum could alter BH retention efficiency in clusters, potentially leading to shorter but more massive merger chains. For further discussion of these effects and the broader impact of IMF variations on a cluster dynamical evolution, \citealp[see][]{Chatterjee_2017, giersz_2019_imf, haghi_2020, weatherford_2021}.

Nonetheless, it is interesting to point out that our estimated rates are broadly consistent with the LVK-inferred rate, $\mathcal{R}_{\rm LVK} = 0.08^{+0.19}_{-0.07}$ yr$^{-1}$ Gpc$^{-3}$. This proof-of-concept study therefore supports a dynamical origin for GW231123. We plan to investigate the effects of different metallicity distributions and IMFs in a follow-up work.



\section{Conclusion}
\label{sec:conclusions}
The discovery of GW231123 challenges current models for BBH formation. This is the first GW-event featuring two highly-spinning massive BHs, possibly representing the first detected IMBH binary, with masses of $m_1 = 137^{+22}_{-17}\msun$, and $m_2 = 103^{+20}_{-52}\msun$. Such masses reside in the upper-mass gap and are precluded by standard models for single stellar evolution. The peculiar properties of the event may be recovered considering a dynamical origin in star clusters.
To explore this scenario, we use a suite of 9 simulations of $5\times 10^7$ BHs each conducted with the semi-analytical code \bpop, investigating BBH formation processes in three different star clusters, namely \young, \globular, and \nuclear, at different metallicities, i.e. $Z=0.02, 0.002$, and $Z=0.0002$. In all simulations we assume a fraction $f_{\rm bin} = 0.6$ of BHs produced in primordial stellar binaries. 
Our main results can be summarised as follows:
\begin{itemize}
\item We find that the formation of GW231123 analogs is possible only at subsolar metallicities ($Z \leq 0.002)$. Regardless of the explored formation scenarios, we find no GW231123-like mergers in environments with solar metallicity (Figure \ref{fig::fiducial_grid}).
\item These systems form either via dynamical pairing and single mergers of BHs in the upper-mass gap (BSE), which originated from primordial stellar binaries evolution, or from chains of hierarchical BBH mergers (H or MC). The first formation process is predominant in low-density clusters (\young, and \globular), while the H-, and MC-channels are suppressed in clusters with such low escape velocities. Conversely, denser environments, such as \nuclear, favor the hierarchical assembly of GW231123-like mergers (Figure \ref{fig::fiducial_grid_Ngen}).
\item We explore the impact of natal spins on the properties of BSE- and H-BBHs by adopting three prescriptions: a low-spin Maxwellian with $\sigma_\chi=0.2$, and two Gaussian distributions with median spins $\chi_\mathrm{med}=0.5$ (medium-spin) and $\chi_\mathrm{med}=0.9$ (high-spin).
We find that higher natal spins enhance the number of GW231123-like BSE-BBHs up to 3 order of magnitude. Higher spins, however, suppress the hierarchical growth of BHs born in the upper-mass gap (H-BBHs) since they generally result in larger post-merger relativistic kicks. Nevertheless, this does not affect strongly the overall number of GW231123-like systems in \nuclear since most of these systems result from chains involving only BHs below the upper-mass gap, i.e. MC-BBH mergers (Figures \ref{fig3:corner}-\ref{fig4:corner} and Table \ref{tab:bbh_origins}).
\item Most simulated GW231123-like mergers exhibit delay times $\sim 0.1-1$ Gyr, indicating that such systems could have formed at redshift $z<1$ in metal-poor environments (Figure \ref{fig:delaytimes}). 
\item We argue that dwarf galaxies at low redshifts, or GCs in Milky Way-like environments, could represent promising hosts and, based on local abundances of dwarf galaxies and of their clusters, we estimate an upper limit to the event rate of the order of $(0.04-36)\times10^{-2} \,\mathrm{yr}^{-1}$ Gpc$^{-3}$ for dwarf galaxies and up to $0.72\,\mathrm{yr}^{-1}$ Gpc$^{-3}$ for GCs in Milky Way-like galaxies (Eqns \eqref{eq3:rate} and \eqref{eq4:rate}). 
\end{itemize}

Our results highlight the impact of stellar properties, such as stellar mergers, in sculpting the mass spectrum of BHs in the upper-mass gap that participate in the build-up of massive BBH mergers like GW231123.

\begin{acknowledgments}
The authors thank Juan Calderón Bustillo and the anonymous referee for the insightful comments and suggestions.
MAS acknowledges funding from the European Union’s Horizon 2020 research and innovation programme under the Marie Skłodowska-Curie grant agreement No.~101025436 (project GRACE-BH) and from the MERAC Foundation through the 2023 MERAC prize. MAS and MB acknowledge the ACME project which has received funding from the European Union's Horizon Europe Research and Innovation programme under Grant Agreement No.~101131928.
MS acknowledges financial support from Large Grant INAF 2024 “Envisioning Tomorrow: prospects and challenges for multi-messenger astronomy in the era of Rubin and Einstein Telescope”, from Fondazione ICSC, Spoke 3 Astrophysics and Cosmos Observations, National Recovery and Resilience Plan (Piano Nazionale di Ripresa e
Resilienza, PNRR) Project ID CN\_00000013 “Italian Research Center on High-Performance Computing, Big Data and Quantum Computing” funded by MUR Missione 4 Componente 2 Investimento 1.4: Potenziamento strutture di ricerca e creazione di “campioni nazionali di R\&S (M4C2-19 )” - Next Generation EU (NGEU), and from the program “Data Science methods for Multi-Messenger Astrophysics \& Multi-Survey Cosmology” funded by the Italian Ministry of University and Research, Programmazione triennale 2021/2023 (DM n.2503 dd. 09/12/2019), Programma Congiunto Scuole.
\end{acknowledgments}

\begin{contribution}
MAS is the main developer and maintainer of the \bpop code. They devised the main topics of this letter and the simulation grid. 
LP performed all the simulations presented here and conducted data analysis and post-processing, producing the main plots.
CU generated all \textsc{SEvN} catalogs, taking care of the initial condition generation, the run of all models, and the conversion from \textsc{SEvN} output tables to the format requested from \bpop. MS is among the core developers of the \textsc{SEvN} population synthesis code and contributed to generate the \textsc{SEvN} catalogs.
MAS, LP, and CU contributed equally to upgrade the \bpop code in order to generate the results presented in this work, and contributed equally to the preparation of this letter. All the authors contributed equally to the interpretation of the results and the finalization of the work in terms of reviewing and writing tasks.
\end{contribution}

\facilities{Simulations are ran on the MERAC2A high-performance computing workstation, hosted at GSSI and financially supported through the MERAC 2023 Prize awarded by the European Astronomical Society and the MERAC Foundation.}

\software{The population synthesis code \bpop is available under reasonable request to the authors. The authors will release the catalogs used to produce the main results of this work on Zenodo, upon publication. 
}

\appendix
\section{BH mass spectrum from primordial stellar binaries}
\label{app:gapdependence}

In this appendix, we explore how different metallicity values affect the formation efficiency of GW231123-like systems in our models. In the following, we focus on metallicities $Z = 0.002$ and $Z = 0.0002$. According to the adopted prescriptions, the maximum BH mass attainable in our models through single stellar evolution is $\sim 52 \ M_\odot$ for $Z = 0.002$ and $\sim 84 \ M_\odot$ for $Z = 0.0002$, i.e. below the masses of GW231123 components. However, the inclusion of stellar mergers and binary evolution substantially overcomes these thresholds. 
\begin{figure}[h!]
    \centering
    \includegraphics[width=0.5\linewidth]{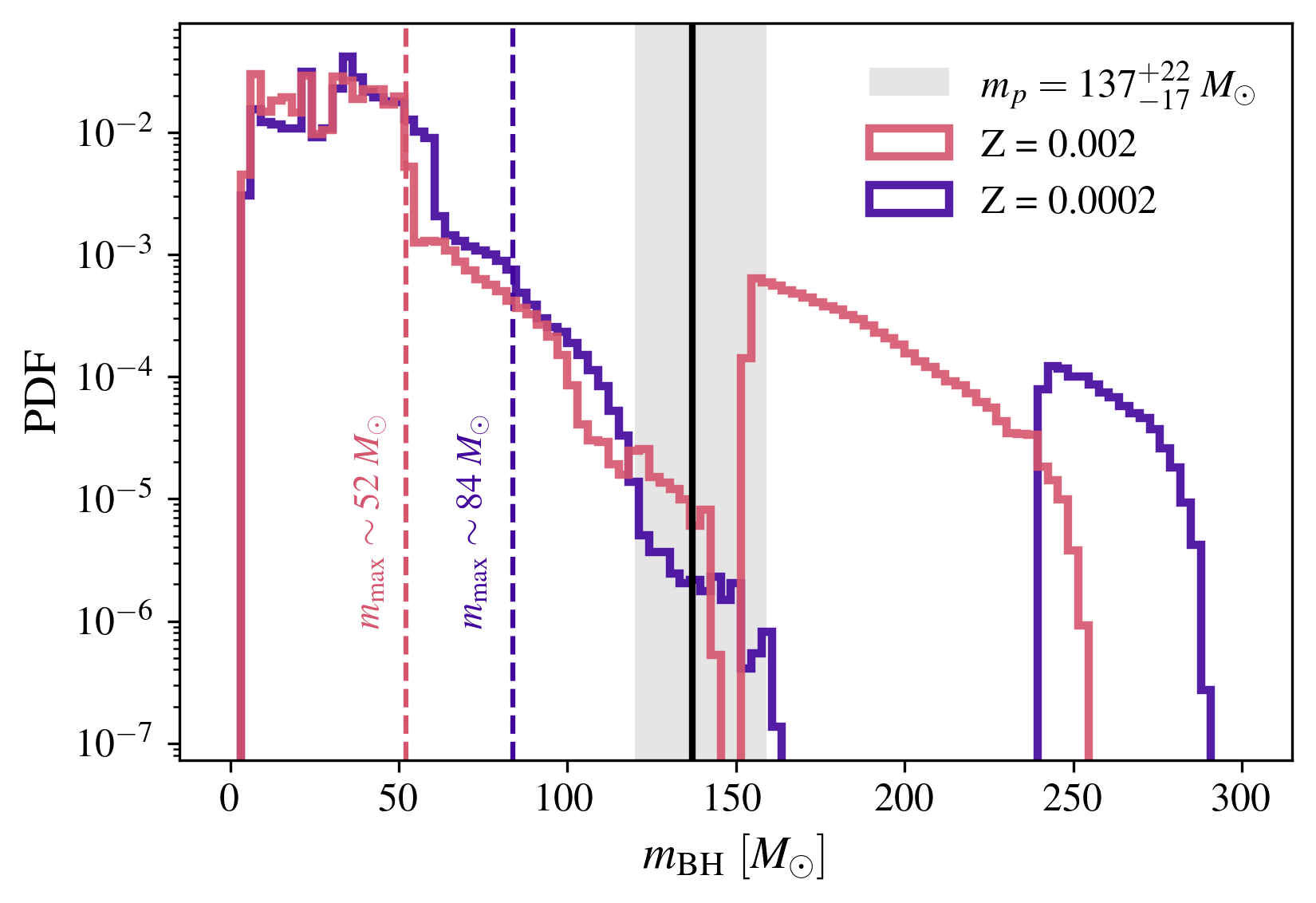}
    \caption{Mass spectrum of BHs produced in primordial stellar binaries for $Z = 0.002$ and $Z = 0.0002$. The black line denotes the primary mass of GW231123 while the gray band indicates the 90 \% C.I.. The dashed lines show the maximum mass that a BH can attain as a product of single stellar evolution.}
    \label{fig:bhmassspectrum}
\end{figure}
In Figure \ref{fig:bhmassspectrum} we compare the mass distribution of BHs produced in stellar binaries and the observational constraints on GW231123 primary mass, $m_p$.  
Although binary processes produce a significant amount of light IMBHs, GW231123 primary mass lies in a region of the spectrum scarcely populated due to PISNe. However, for $Z = 0.002$, the upper mass limit of the PISN gap falls just inside the 90$\%$ C.I. of GW231123 primary mass, thus significantly enhancing the probability to find similar BHs in our models. This peculiarity highlights two crucial aspects that can affect the astrophysical interpretation of sources like GW231123.

First, the assumptions on the PISN mechanism and primordial binary properties in star clusters can significantly shift the boundaries of the PISN gap, thereby crucially affecting the likelihood of producing GW231123 analogs. Second, improving the accuracy and precision of the inferred parameters of such GW sources can be crucial to assess the physical processes and environments leading to their formation. In the case of GW231123, shifting the 90 \% credible interval by $5 \msun$, i.e. assuming the primary to be slightly lighter, reduces the fraction of GW231123-like systems in our analysis by a factor of 1--8. We note, however, that this effect may be mitigated by considering the full distribution around the median of $m_p$, rather than a rigid interval encompassing the 90\% C.I..

\bibliography{biblio}

\begin{thebibliography}{}
\expandafter\ifx\csname natexlab\endcsname\relax\def\natexlab#1{#1}\fi
\providecommand{\url}[1]{\href{#1}{#1}}
\providecommand{\dodoi}[1]{doi:~\href{http://doi.org/#1}{\nolinkurl{#1}}}
\providecommand{\doeprint}[1]{\href{http://ascl.net/#1}{\nolinkurl{http://ascl.net/#1}}}
\providecommand{\doarXiv}[1]{\href{https://arxiv.org/abs/#1}{\nolinkurl{https://arxiv.org/abs/#1}}}

\bibitem[{P. Ajith {et~al.}(2011)Ajith, Hannam, Husa, Chen, Br\"ugmann, Dorband, M\"uller, Ohme, Pollney, Reisswig, Santamar\'{\i}a, \& Seiler}]{Ajith2011}
Ajith, P., Hannam, M., Husa, S., {et~al.} 2011, \bibinfo{title}{Inspiral-Merger-Ringdown Waveforms for Black-Hole Binaries with Nonprecessing Spins,} Phys. Rev. Lett., 106, 241101, \dodoi{10.1103/PhysRevLett.106.241101}

\bibitem[{F. {Antonini} {et~al.}(2019){Antonini}, {Gieles}, \& {Gualandris}}]{Antonini_2019}
{Antonini}, F., {Gieles}, M., \& {Gualandris}, A. 2019, \bibinfo{title}{{Black hole growth through hierarchical black hole mergers in dense star clusters: implications for gravitational wave detections},} \mnras, 486, 5008, \dodoi{10.1093/mnras/stz1149}

\bibitem[{F. Antonini \& F.~A. Rasio(2016)Antonini \& Rasio}]{Antonini_2016}
Antonini, F., \& Rasio, F.~A. 2016, \bibinfo{title}{MERGING BLACK HOLE BINARIES IN GALACTIC NUCLEI: IMPLICATIONS FOR ADVANCED-LIGO DETECTIONS,} The Astrophysical Journal, 831, 187, \dodoi{10.3847/0004-637x/831/2/187}

\bibitem[{C. {Ara{\'u}jo-{\'A}lvarez} {et~al.}(2024){Ara{\'u}jo-{\'A}lvarez}, {Wong}, {Liu}, \& {Calder{\'o}n Bustillo}}]{alvarez_2024}
{Ara{\'u}jo-{\'A}lvarez}, C., {Wong}, H. W.~Y., {Liu}, A., \& {Calder{\'o}n Bustillo}, J. 2024, \bibinfo{title}{{Kicking Time Back in Black Hole Mergers: Ancestral Masses, Spins, Birth Recoils, and Hierarchical-formation Viability of GW190521},} \apj, 977, 220, \dodoi{10.3847/1538-4357/ad90a9}

\bibitem[{M. {Arca Sedda} \& M. {Benacquista}(2019){Arca Sedda} \& {Benacquista}}]{ArcaSedda_Benacquista_2019}
{Arca Sedda}, M., \& {Benacquista}, M. 2019, \bibinfo{title}{{Using final black hole spins and masses to infer the formation history of the observed population of gravitational wave sources},} \mnras, 482, 2991, \dodoi{10.1093/mnras/sty2764}

\bibitem[{M. {Arca Sedda} {et~al.}(2020){Arca Sedda}, {Gualandris}, {Do}, {Feldmeier-Krause}, {Neumayer}, \& {Erkal}}]{2020ApJ...901L..29A}
{Arca Sedda}, M., {Gualandris}, A., {Do}, T., {et~al.} 2020, \bibinfo{title}{{On the Origin of a Rotating Metal-poor Stellar Population in the Milky Way Nuclear Cluster},} \apjl, 901, L29, \dodoi{10.3847/2041-8213/abb245}

\bibitem[{M. {Arca Sedda} {et~al.}(2024{\natexlab{a}}){Arca Sedda}, {Kamlah}, {Spurzem}, {Rizzuto}, {Giersz}, {Naab}, \& {Berczik}}]{AS_DragonII_3}
{Arca Sedda}, M., {Kamlah}, A. W.~H., {Spurzem}, R., {et~al.} 2024{\natexlab{a}}, \bibinfo{title}{{The DRAGON-II simulations - III. Compact binary mergers in clusters with up to 1 million stars: mass, spin, eccentricity, merger rate, and pair instability supernovae rate},} \mnras, 528, 5140, \dodoi{10.1093/mnras/stad3951}

\bibitem[{M. {Arca Sedda} {et~al.}(2023){Arca Sedda}, {Kamlah}, {Spurzem}, {Rizzuto}, {Naab}, {Giersz}, \& {Berczik}}]{AS_DragonII_2}
{Arca Sedda}, M., {Kamlah}, A. W.~H., {Spurzem}, R., {et~al.} 2023, \bibinfo{title}{{The DRAGON-II simulations - II. Formation mechanisms, mass, and spin of intermediate-mass black holes in star clusters with up to 1 million stars},} \mnras, 526, 429, \dodoi{10.1093/mnras/stad2292}

\bibitem[{M. Arca~Sedda {et~al.}(2023)Arca~Sedda, Mapelli, Benacquista, \& Spera}]{Arca_Sedda_2023}
Arca~Sedda, M., Mapelli, M., Benacquista, M., \& Spera, M. 2023, \bibinfo{title}{Isolated and dynamical black hole mergers with<tt>B-POP</tt>: the role of star formation and dynamics, star cluster evolution, natal kicks, mass and spins, and hierarchical mergers,} Monthly Notices of the Royal Astronomical Society, 520, 5259–5282, \dodoi{10.1093/mnras/stad331}

\bibitem[{M. Arca~Sedda {et~al.}(2020)Arca~Sedda, Mapelli, Spera, Benacquista, \& Giacobbo}]{arca_sedda_fingerprints_2020}
Arca~Sedda, M., Mapelli, M., Spera, M., Benacquista, M., \& Giacobbo, N. 2020, \bibinfo{title}{Fingerprints of {Binary} {Black} {Hole} {Formation} {Channels} {Encoded} in the {Mass} and {Spin} of {Merger} {Remnants},} The Astrophysical Journal, 894, 133, \dodoi{10.3847/1538-4357/ab88b2}

\bibitem[{M. {Arca Sedda} {et~al.}(2023){Arca Sedda}, {Naoz}, \& {Kocsis}}]{2023Univ....9..138A}
{Arca Sedda}, M., {Naoz}, S., \& {Kocsis}, B. 2023, \bibinfo{title}{{Quiescent and Active Galactic Nuclei as Factories of Merging Compact Objects in the Era of Gravitational Wave Astronomy},} Universe, 9, 138, \dodoi{10.3390/universe9030138}

\bibitem[{M. {Arca-Sedda} {et~al.}(2021){Arca-Sedda}, {Rizzuto}, {Naab}, {Ostriker}, {Giersz}, \& {Spurzem}}]{arcasedda_2021}
{Arca-Sedda}, M., {Rizzuto}, F.~P., {Naab}, T., {et~al.} 2021, \bibinfo{title}{{Breaching the Limit: Formation of GW190521-like and IMBH Mergers in Young Massive Clusters},} \apj, 920, 128, \dodoi{10.3847/1538-4357/ac1419}

\bibitem[{M. {Arca Sedda} {et~al.}(2024{\natexlab{b}}){Arca Sedda}, {Kamlah}, {Spurzem}, {Giersz}, {Berczik}, {Rastello}, {Iorio}, {Mapelli}, {Gatto}, \& {Grebel}}]{AS_DragonII_1}
{Arca Sedda}, M., {Kamlah}, A. W.~H., {Spurzem}, R., {et~al.} 2024{\natexlab{b}}, \bibinfo{title}{{The DRAGON-II simulations - I. Evolution of single and binary compact objects in star clusters with up to 1 million stars},} \mnras, 528, 5119, \dodoi{10.1093/mnras/stad3952}

\bibitem[{K.~G. Arun {et~al.}(2009)Arun, Buonanno, Faye, \& Ochsner}]{Arun2009}
Arun, K.~G., Buonanno, A., Faye, G., \& Ochsner, E. 2009, \bibinfo{title}{Higher-order spin effects in the amplitude and phase of gravitational waveforms emitted by inspiraling compact binaries: Ready-to-use gravitational waveforms,} Phys. Rev. D, 79, 104023, \dodoi{10.1103/PhysRevD.79.104023}

\bibitem[{A. {Ballone} {et~al.}(2023){Ballone}, {Costa}, {Mapelli}, {MacLeod}, {Torniamenti}, \& {Pacheco-Arias}}]{Ballone2023}
{Ballone}, A., {Costa}, G., {Mapelli}, M., {et~al.} 2023, \bibinfo{title}{{Formation of black holes in the pair-instability mass gap: hydrodynamical simulations of a head-on massive star collision},} \mnras, 519, 5191, \dodoi{10.1093/mnras/stac3752}

\bibitem[{S. Banerjee(2018)Banerjee}]{Banerjee2018}
Banerjee, S. 2018, \bibinfo{title}{Stellar-Mass Black Holes in Young Massive and Open Stellar Clusters and Their Role in Gravitational-Wave Generation - {{II}},} 473, 909, \dodoi{10.1093/mnras/stx2347}

\bibitem[{S. Banerjee {et~al.}(2010)Banerjee, Baumgardt, \& Kroupa}]{banerjeeStellarmassBlackHoles2010}
Banerjee, S., Baumgardt, H., \& Kroupa, P. 2010, \bibinfo{title}{Stellar-Mass Black Holes in Star Clusters: Implications for Gravitational Wave Radiation,} Monthly Notices of the Royal Astronomical Society, 402, 371, \dodoi{10.1111/j.1365-2966.2009.15880.x}

\bibitem[{K. Belczynski {et~al.}(2002)Belczynski, Kalogera, \& Bulik}]{belczynski_comprehensive_2002}
Belczynski, K., Kalogera, V., \& Bulik, T. 2002, \bibinfo{title}{A {Comprehensive} {Study} of {Binary} {Compact} {Objects} as {Gravitational} {Wave} {Sources}: {Evolutionary} {Channels}, {Rates}, and {Physical} {Properties},} The Astrophysical Journal, 572, 407, \dodoi{10.1086/340304}

\bibitem[{E. {Berti} {et~al.}(2007){Berti}, {Cardoso}, {Gonzalez}, {Sperhake}, {Hannam}, {Husa}, \& {Br{\"u}gmann}}]{Berti2007}
{Berti}, E., {Cardoso}, V., {Gonzalez}, J.~A., {et~al.} 2007, \bibinfo{title}{{Inspiral, merger, and ringdown of unequal mass black hole binaries: A multipolar analysis},} \prd, 76, 064034, \dodoi{10.1103/PhysRevD.76.064034}

\bibitem[{J.~P. {Brodie} \& J. {Strader}(2006){Brodie} \& {Strader}}]{2006ARA&A..44..193B}
{Brodie}, J.~P., \& {Strader}, J. 2006, \bibinfo{title}{{Extragalactic Globular Clusters and Galaxy Formation},} \araa, 44, 193, \dodoi{10.1146/annurev.astro.44.051905.092441}

\bibitem[{J.~C. {Bustillo} {et~al.}(2021){Bustillo}, {Sanchis-Gual}, {Torres-Forn{\'e}}, \& {Font}}]{ecc_bustillo_2021}
{Bustillo}, J.~C., {Sanchis-Gual}, N., {Torres-Forn{\'e}}, A., \& {Font}, J.~A. 2021, \bibinfo{title}{{Confusing Head-On Collisions with Precessing Intermediate-Mass Binary Black Hole Mergers},} \prl, 126, 201101, \dodoi{10.1103/PhysRevLett.126.201101}

\bibitem[{A. {Calabr{\`o}} {et~al.}(2017){Calabr{\`o}}, {Amor{\'\i}n}, {Fontana}, {P{\'e}rez-Montero}, {Lemaux}, {Ribeiro}, {Bardelli}, {Castellano}, {Contini}, {De Barros}, {Garilli}, {Grazian}, {Guaita}, {Hathi}, {Koekemoer}, {Le F{\`e}vre}, {Maccagni}, {Pentericci}, {Schaerer}, {Talia}, {Tasca}, \& {Zucca}}]{2017A&A...601A..95C}
{Calabr{\`o}}, A., {Amor{\'\i}n}, R., {Fontana}, A., {et~al.} 2017, \bibinfo{title}{{Characterization of star-forming dwarf galaxies at 0.1 {\ensuremath{\lesssim}}z {\ensuremath{\lesssim}} 0.9 in VUDS: probing the low-mass end of the mass-metallicity relation},} \aap, 601, A95, \dodoi{10.1051/0004-6361/201629762}

\bibitem[{M. {Campanelli} {et~al.}(2007){Campanelli}, {Lousto}, {Zlochower}, \& {Merritt}}]{campanelli_2007}
{Campanelli}, M., {Lousto}, C.~O., {Zlochower}, Y., \& {Merritt}, D. 2007, \bibinfo{title}{{Maximum Gravitational Recoil},} \prl, 98, 231102, \dodoi{10.1103/PhysRevLett.98.231102}

\bibitem[{S. {Chatterjee} {et~al.}(2017){Chatterjee}, {Rodriguez}, \& {Rasio}}]{Chatterjee_2017}
{Chatterjee}, S., {Rodriguez}, C.~L., \& {Rasio}, F.~A. 2017, \bibinfo{title}{{Binary Black Holes in Dense Star Clusters: Exploring the Theoretical Uncertainties},} \apj, 834, 68, \dodoi{10.3847/1538-4357/834/1/68}

\bibitem[{T.~L.~S. Collaboration {et~al.}(2022)Collaboration, {the Virgo Collaboration}, \& {the KAGRA Collaboration}}]{theligoscientificcollaborationPopulationMergingCompact2022}
Collaboration, T. L.~S., {the Virgo Collaboration}, \& {the KAGRA Collaboration}. 2022, The Population of Merging Compact Binaries Inferred Using Gravitational Waves through {{GWTC-3}}, arXiv.
\newblock \doarXiv{2111.03634}

\bibitem[{G. Costa {et~al.}(2022)Costa, Ballone, Mapelli, \& Bressan}]{Costa2022}
Costa, G., Ballone, A., Mapelli, M., \& Bressan, A. 2022, \bibinfo{title}{Formation of black holes in the pair-instability mass gap: Evolution of a post-collision star,} Monthly Notices of the Royal Astronomical Society, 516, 1072, \dodoi{10.1093/mnras/stac2222}

\bibitem[{G. Costa {et~al.}(2021)Costa, Bressan, Mapelli, Marigo, Iorio, \& Spera}]{costaFormationGW190521Stellar2021}
Costa, G., Bressan, A., Mapelli, M., {et~al.} 2021, \bibinfo{title}{Formation of {{GW190521}} from Stellar Evolution: The Impact of the Hydrogen-Rich Envelope, Dredge-up, and {$^{12}C$}({$\alpha$}, {$\gamma$}){$^{16}O$} Rate on the Pair-Instability Black Hole Mass Gap,} Monthly Notices of the Royal Astronomical Society, 501, 4514, \dodoi{10.1093/mnras/staa3916}

\bibitem[{G. Costa {et~al.}(2023)Costa, Mapelli, Iorio, Santoliquido, Escobar, Klessen, \& Bressan}]{costaMassiveBinaryBlack2023}
Costa, G., Mapelli, M., Iorio, G., {et~al.} 2023, Massive Binary Black Holes from {{Population II}} and {{III}} Stars, arXiv.
\newblock \doarXiv{2303.15511}

\bibitem[{M. Curti {et~al.}(2019)Curti, Mannucci, Cresci, \& Maiolino}]{Curti2020}
Curti, M., Mannucci, F., Cresci, G., \& Maiolino, R. 2019, \bibinfo{title}{The mass–metallicity and the fundamental metallicity relation revisited on a fully Te-based abundance scale for galaxies,} Monthly Notices of the Royal Astronomical Society, 491, 944, \dodoi{10.1093/mnras/stz2910}

\bibitem[{U.~N. Di~Carlo {et~al.}(2019)Di~Carlo, Giacobbo, Mapelli, Pasquato, Spera, Wang, \& Haardt}]{dicarloMergingBlackHoles2019a}
Di~Carlo, U.~N., Giacobbo, N., Mapelli, M., {et~al.} 2019, \bibinfo{title}{Merging Black Holes in Young Star Clusters,} Monthly Notices of the Royal Astronomical Society, 487, 2947, \dodoi{10.1093/mnras/stz1453}

\bibitem[{U.~N. Di~Carlo {et~al.}(2021)Di~Carlo, Mapelli, Pasquato, Rastello, Ballone, Dall'Amico, Giacobbo, Iorio, Spera, Torniamenti, \& Haardt}]{DiCarlo2021a}
Di~Carlo, U.~N., Mapelli, M., Pasquato, M., {et~al.} 2021, \bibinfo{title}{Intermediate-Mass Black Holes from Stellar Mergers in Young Star Clusters,} 507, 5132, \dodoi{10.1093/mnras/stab2390}

\bibitem[{T. {Do} {et~al.}(2020){Do}, {David Martinez}, {Kerzendorf}, {Feldmeier-Krause}, {Arca Sedda}, {Neumayer}, \& {Gualandris}}]{2020ApJ...901L..28D}
{Do}, T., {David Martinez}, G., {Kerzendorf}, W., {et~al.} 2020, \bibinfo{title}{{Revealing the Formation of the Milky Way Nuclear Star Cluster via Chemo-dynamical Modeling},} \apjl, 901, L28, \dodoi{10.3847/2041-8213/abb246}

\bibitem[{R. Farmer {et~al.}(2019)Farmer, Renzo, {de Mink}, Marchant, \& Justham}]{Farmer2019}
Farmer, R., Renzo, M., {de Mink}, S.~E., Marchant, P., \& Justham, S. 2019, \bibinfo{title}{Mind the Gap: {{The}} Location of the Lower Edge of the Pair-Instability Supernova Black Hole Mass Gap,} 887, 53, \dodoi{10.3847/1538-4357/ab518b}

\bibitem[{C.~L. Fryer {et~al.}(2012)Fryer, Belczynski, Wiktorowicz, Dominik, Kalogera, \& Holz}]{fryer_compact_2012}
Fryer, C.~L., Belczynski, K., Wiktorowicz, G., {et~al.} 2012, \bibinfo{title}{Compact Remnant Mass Function: {{Dependence}} on the Explosion Mechanism and Metallicity,} 749, 91, \dodoi{10.1088/0004-637X/749/1/91}

\bibitem[{J. {Fuller} \& L. {Ma}(2019){Fuller} \& {Ma}}]{Fuller2019}
{Fuller}, J., \& {Ma}, L. 2019, \bibinfo{title}{{Most Black Holes Are Born Very Slowly Rotating},} \apjl, 881, L1, \dodoi{10.3847/2041-8213/ab339b}

\bibitem[{V. {Gayathri} {et~al.}(2020){Gayathri}, {Healy}, {Lange}, {O'Brien}, {Szczepanczyk}, {Bartos}, {Campanelli}, {Klimenko}, {Lousto}, \& {O'Shaughnessy}}]{ecc_gayathri_2020}
{Gayathri}, V., {Healy}, J., {Lange}, J., {et~al.} 2020, \bibinfo{title}{{Eccentricity Estimate for Black Hole Mergers with Numerical Relativity Simulations},} arXiv e-prints, arXiv:2009.05461, \dodoi{10.48550/arXiv.2009.05461}

\bibitem[{M. {Giersz} {et~al.}(2019){Giersz}, {Askar}, {Wang}, {Hypki}, {Leveque}, \& {Spurzem}}]{giersz_2019_imf}
{Giersz}, M., {Askar}, A., {Wang}, L., {et~al.} 2019, \bibinfo{title}{{MOCCA survey data base- I. Dissolution of tidally filling star clusters harbouring black hole subsystems},} \mnras, 487, 2412, \dodoi{10.1093/mnras/stz1460}

\bibitem[{J.~A. {Gonz{\'a}lez} {et~al.}(2007){Gonz{\'a}lez}, {Hannam}, {Sperhake}, {Br{\"u}gmann}, \& {Husa}}]{gonzalez_2007}
{Gonz{\'a}lez}, J.~A., {Hannam}, M., {Sperhake}, U., {Br{\"u}gmann}, B., \& {Husa}, S. 2007, \bibinfo{title}{{Supermassive Recoil Velocities for Binary Black-Hole Mergers with Antialigned Spins},} \prl, 98, 231101, \dodoi{10.1103/PhysRevLett.98.231101}

\bibitem[{H. {Haghi} {et~al.}(2020){Haghi}, {Safaei}, {Zonoozi}, \& {Kroupa}}]{haghi_2020}
{Haghi}, H., {Safaei}, G., {Zonoozi}, A.~H., \& {Kroupa}, P. 2020, \bibinfo{title}{{The Lifetimes of Star Clusters Born with a Top-heavy IMF},} \apj, 904, 43, \dodoi{10.3847/1538-4357/abbfb0}

\bibitem[{W.~E. {Harris}(1996){Harris}}]{1996AJ....112.1487H}
{Harris}, W.~E. 1996, \bibinfo{title}{{A Catalog of Parameters for Globular Clusters in the Milky Way},} \aj, 112, 1487, \dodoi{10.1086/118116}

\bibitem[{W.~E. {Harris}(2010){Harris}}]{2010arXiv1012.3224H}
{Harris}, W.~E. 2010, \bibinfo{title}{{A New Catalog of Globular Clusters in the Milky Way},} arXiv e-prints, arXiv:1012.3224, \dodoi{10.48550/arXiv.1012.3224}

\bibitem[{A. Heger \& S.~E. Woosley(2002)Heger \& Woosley}]{Heger2002}
Heger, A., \& Woosley, S.~E. 2002, \bibinfo{title}{The Nucleosynthetic Signature of Population {{III}},} 567, 532, \dodoi{10.1086/338487}

\bibitem[{D.~C. Heggie(1975)Heggie}]{Heggie1975}
Heggie, D.~C. 1975, \bibinfo{title}{Binary Evolution in Stellar Dynamics.,} 173, 729, \dodoi{10.1093/mnras/173.3.729}

\bibitem[{D.~D. Hendriks {et~al.}(2023)Hendriks, {van Son}, Renzo, Izzard, \& Farmer}]{hendriksPulsationalPairinstabilitySupernovae2023}
Hendriks, D.~D., {van Son}, L. A.~C., Renzo, M., Izzard, R.~G., \& Farmer, R. 2023, Pulsational Pair-Instability Supernovae in Gravitational-Wave and Electromagnetic Transients, arXiv.
\newblock \doarXiv{2309.09339}

\bibitem[{F. {Hofmann} {et~al.}(2016){Hofmann}, {Barausse}, \& {Rezzolla}}]{Hofmann2016}
{Hofmann}, F., {Barausse}, E., \& {Rezzolla}, L. 2016, \bibinfo{title}{{The Final Spin from Binary Black Holes in Quasi-circular Orbits},} \apjl, 825, L19, \dodoi{10.3847/2041-8205/825/2/L19}

\bibitem[{S.~A. {Hughes} \& R.~D. {Blandford}(2003){Hughes} \& {Blandford}}]{2003ApJ...585L.101H}
{Hughes}, S.~A., \& {Blandford}, R.~D. 2003, \bibinfo{title}{{Black Hole Mass and Spin Coevolution by Mergers},} \apjl, 585, L101, \dodoi{10.1086/375495}

\bibitem[{G. Iorio {et~al.}(2023)Iorio, Mapelli, Costa, Spera, Escobar, Sgalletta, Trani, Korb, Santoliquido, Dall'Amico, Gaspari, \& Bressan}]{iorioCompactObjectMergers2023}
Iorio, G., Mapelli, M., Costa, G., {et~al.} 2023, \bibinfo{title}{Compact Object Mergers: Exploring Uncertainties from Stellar and Binary Evolution with {{SEVN}},} Monthly Notices of the Royal Astronomical Society, 524, 426, \dodoi{10.1093/mnras/stad1630}

\bibitem[{N.~V. {Kharchenko} {et~al.}(2013){Kharchenko}, {Piskunov}, {Schilbach}, {R{\"o}ser}, \& {Scholz}}]{2013A&A...558A..53K}
{Kharchenko}, N.~V., {Piskunov}, A.~E., {Schilbach}, E., {R{\"o}ser}, S., \& {Scholz}, R.~D. 2013, \bibinfo{title}{{Global survey of star clusters in the Milky Way. II. The catalogue of basic parameters},} \aap, 558, A53, \dodoi{10.1051/0004-6361/201322302}

\bibitem[{E.~N. {Kirby} {et~al.}(2013){Kirby}, {Cohen}, {Guhathakurta}, {Cheng}, {Bullock}, \& {Gallazzi}}]{2013ApJ...779..102K}
{Kirby}, E.~N., {Cohen}, J.~G., {Guhathakurta}, P., {et~al.} 2013, \bibinfo{title}{{The Universal Stellar Mass-Stellar Metallicity Relation for Dwarf Galaxies},} \apj, 779, 102, \dodoi{10.1088/0004-637X/779/2/102}

\bibitem[{R.~K. {Kopparapu} {et~al.}(2008){Kopparapu}, {Hanna}, {Kalogera}, {O'Shaughnessy}, {Gonz{\'a}lez}, {Brady}, \& {Fairhurst}}]{2008ApJ...675.1459K}
{Kopparapu}, R.~K., {Hanna}, C., {Kalogera}, V., {et~al.} 2008, \bibinfo{title}{{Host Galaxies Catalog Used in LIGO Searches for Compact Binary Coalescence Events},} \apj, 675, 1459, \dodoi{10.1086/527348}

\bibitem[{K. {Kremer} {et~al.}(2020){Kremer}, {Spera}, {Becker}, {Chatterjee}, {Di Carlo}, {Fragione}, {Rodriguez}, {Ye}, \& {Rasio}}]{Kremer2020}
{Kremer}, K., {Spera}, M., {Becker}, D., {et~al.} 2020, \bibinfo{title}{{Populating the Upper Black Hole Mass Gap through Stellar Collisions in Young Star Clusters},} The Astrophysical Journal, 903, 45, \dodoi{10.3847/1538-4357/abb945}

\bibitem[{K. {Kritos} {et~al.}(2024){Kritos}, {Strokov}, {Baibhav}, \& {Berti}}]{Kritos2024}
{Kritos}, K., {Strokov}, V., {Baibhav}, V., \& {Berti}, E. 2024, \bibinfo{title}{{Dynamical formation of black hole binaries in dense star clusters: Rapid cluster evolution code},} \prd, 110, 043023, \dodoi{10.1103/PhysRevD.110.043023}

\bibitem[{P. Kroupa(2001)Kroupa}]{Kroupa2001}
Kroupa, P. 2001, \bibinfo{title}{On the Variation of the Initial Mass Function,} 322, 231, \dodoi{10.1046/j.1365-8711.2001.04022.x}

\bibitem[{S.~S. {Larsen}(2009){Larsen}}]{larsen_2009}
{Larsen}, S.~S. 2009, \bibinfo{title}{{The mass function of young star clusters in spiral galaxies},} \aap, 503, 467, \dodoi{10.1051/0004-6361/200811212e}

\bibitem[{M. Limongi \& A. Chieffi(2018)Limongi \& Chieffi}]{limongi_presupernova_2018}
Limongi, M., \& Chieffi, A. 2018, \bibinfo{title}{Presupernova Evolution and Explosive Nucleosynthesis of Rotating Massive Stars in the Metallicity Range -3 {$\leq$} [Fe/{{H}}] {$\leq$} 0,} 237, 13, \dodoi{10.3847/1538-4365/aacb24}

\bibitem[{C.~O. {Lousto} \& Y. {Zlochower}(2008){Lousto} \& {Zlochower}}]{lousto_2008}
{Lousto}, C.~O., \& {Zlochower}, Y. 2008, \bibinfo{title}{{Further insight into gravitational recoil},} \prd, 77, 044028, \dodoi{10.1103/PhysRevD.77.044028}

\bibitem[{C.~O. {Lousto} {et~al.}(2012){Lousto}, {Zlochower}, {Dotti}, \& {Volonteri}}]{lousto_2012}
{Lousto}, C.~O., {Zlochower}, Y., {Dotti}, M., \& {Volonteri}, M. 2012, \bibinfo{title}{{Gravitational recoil from accretion-aligned black-hole binaries},} \prd, 85, 084015, \dodoi{10.1103/PhysRevD.85.084015}

\bibitem[{X. {Ma} {et~al.}(2016){Ma}, {Hopkins}, {Faucher-Gigu{\`e}re}, {Zolman}, {Muratov}, {Kere{\v{s}}}, \& {Quataert}}]{2016MNRAS.456.2140M}
{Ma}, X., {Hopkins}, P.~F., {Faucher-Gigu{\`e}re}, C.-A., {et~al.} 2016, \bibinfo{title}{{The origin and evolution of the galaxy mass-metallicity relation},} \mnras, 456, 2140, \dodoi{10.1093/mnras/stv2659}

\bibitem[{P. {Mahapatra} {et~al.}(2024){Mahapatra}, {Chattopadhyay}, {Gupta}, {Antonini}, {Favata}, {Sathyaprakash}, \& {Arun}}]{mahapatra_2024}
{Mahapatra}, P., {Chattopadhyay}, D., {Gupta}, A., {et~al.} 2024, \bibinfo{title}{{Reconstructing the Genealogy of LIGO-Virgo Black Holes},} \apj, 975, 117, \dodoi{10.3847/1538-4357/ad781b}

\bibitem[{F. Mannucci {et~al.}(2010)Mannucci, Cresci, Maiolino, Marconi, \& Gnerucci}]{Mannucci2010}
Mannucci, F., Cresci, G., Maiolino, R., Marconi, A., \& Gnerucci, A. 2010, \bibinfo{title}{A fundamental relation between mass, star formation rate and metallicity in local and high-redshift galaxies,} Monthly Notices of the Royal Astronomical Society, 408, 2115, \dodoi{10.1111/j.1365-2966.2010.17291.x}

\bibitem[{M. Mapelli(2021{\natexlab{a}})Mapelli}]{mapelliFormationChannelsSingle2021a}
Mapelli, M. 2021{\natexlab{a}}, \bibinfo{title}{Formation {{Channels}} of {{Single}} and {{Binary Stellar-Mass Black Holes}},} in Handbook of {{Gravitational Wave Astronomy}}, 16, \dodoi{10.1007/978-981-15-4702-7_16-1}

\bibitem[{M. Mapelli(2021{\natexlab{b}})Mapelli}]{Mapelli2021}
Mapelli, M. 2021{\natexlab{b}}, \bibinfo{title}{Formation Channels of Single and Binary Stellar-Mass Black Holes,} in Handbook of Gravitational Wave Astronomy No.~16, 16, \dodoi{10.1007/978-981-15-4702-7_16-1}

\bibitem[{M. Mapelli {et~al.}(2022)Mapelli, Bouffanais, Santoliquido, Arca~Sedda, \& Artale}]{mapelliCosmicEvolutionBinary2022a}
Mapelli, M., Bouffanais, Y., Santoliquido, F., Arca~Sedda, M., \& Artale, M.~C. 2022, \bibinfo{title}{The Cosmic Evolution of Binary Black Holes in Young, Globular, and Nuclear Star Clusters: Rates, Masses, Spins, and Mixing Fractions,} Monthly Notices of the Royal Astronomical Society, 511, 5797, \dodoi{10.1093/mnras/stac422}

\bibitem[{P. Marchant {et~al.}(2019)Marchant, Renzo, Farmer, Pappas, Taam, De~Mink, \& Kalogera}]{marchantPulsationalPairinstabilitySupernovae2019a}
Marchant, P., Renzo, M., Farmer, R., {et~al.} 2019, \bibinfo{title}{Pulsational {{Pair-instability Supernovae}} in {{Very Close Binaries}},} ApJ, 882, 36, \dodoi{10.3847/1538-4357/ab3426}

\bibitem[{M. {Marks} {et~al.}(2012){Marks}, {Kroupa}, {Dabringhausen}, \& {Pawlowski}}]{marks_2012}
{Marks}, M., {Kroupa}, P., {Dabringhausen}, J., \& {Pawlowski}, M.~S. 2012, \bibinfo{title}{{Evidence for top-heavy stellar initial mass functions with increasing density and decreasing metallicity},} \mnras, 422, 2246, \dodoi{10.1111/j.1365-2966.2012.20767.x}

\bibitem[{B. {McKernan} {et~al.}(2012){McKernan}, {Ford}, {Lyra}, \& {Perets}}]{2012MNRAS.425..460M}
{McKernan}, B., {Ford}, K.~E.~S., {Lyra}, W., \& {Perets}, H.~B. 2012, \bibinfo{title}{{Intermediate mass black holes in AGN discs - I. Production and growth},} \mnras, 425, 460, \dodoi{10.1111/j.1365-2966.2012.21486.x}

\bibitem[{M.~C. {Miller}(2002){Miller}}]{Miller2002}
{Miller}, M.~C. 2002, \bibinfo{title}{{Gravitational Radiation from Intermediate-Mass Black Holes},} \apj, 581, 438, \dodoi{10.1086/344156}

\bibitem[{M.~C. {Miller} \& D.~P. {Hamilton}(2002){Miller} \& {Hamilton}}]{Miller_IMBH_2002}
{Miller}, M.~C., \& {Hamilton}, D.~P. 2002, \bibinfo{title}{{Production of intermediate-mass black holes in globular clusters},} \mnras, 330, 232, \dodoi{10.1046/j.1365-8711.2002.05112.x}

\bibitem[{M.~C. Miller \& D.~P. Hamilton(2002)Miller \& Hamilton}]{miller_four-body_2002}
Miller, M.~C., \& Hamilton, D.~P. 2002, \bibinfo{title}{Four-{Body} {Effects} in {Globular} {Cluster} {Black} {Hole} {Coalescence},} The Astrophysical Journal, 576, 894, \dodoi{10.1086/341788}

\bibitem[{D. {Minniti} {et~al.}(2017){Minniti}, {Geisler}, {Alonso-Garc{\'\i}a}, {Palma}, {Beam{\'\i}n}, {Borissova}, {Catelan}, {Clari{\'a}}, {Cohen}, {Contreras Ramos}, {Dias}, {Fern{\'a}ndez-Trincado}, {G{\'o}mez}, {Hempel}, {Ivanov}, {Kurtev}, {Lucas}, {Moni-Bidin}, {Pullen}, {Ram{\'\i}rez Alegr{\'\i}a}, {Saito}, \& {Valenti}}]{2017ApJ...849L..24M}
{Minniti}, D., {Geisler}, D., {Alonso-Garc{\'\i}a}, J., {et~al.} 2017, \bibinfo{title}{{New VVV Survey Globular Cluster Candidates in the Milky Way Bulge},} \apjl, 849, L24, \dodoi{10.3847/2041-8213/aa95b8}

\bibitem[{M. Moe \& R.~D. Stefano(2017)Moe \& Stefano}]{Moe_2017}
Moe, M., \& Stefano, R.~D. 2017, \bibinfo{title}{Mind Your Ps and Qs: The Interrelation between Period (P) and Mass-ratio (Q) Distributions of Binary Stars,} The Astrophysical Journal Supplement Series, 230, 15, \dodoi{10.3847/1538-4365/aa6fb6}

\bibitem[{N. {Neumayer} {et~al.}(2020){Neumayer}, {Seth}, \& {B{\"o}ker}}]{2020A&ARv..28....4N}
{Neumayer}, N., {Seth}, A., \& {B{\"o}ker}, T. 2020, \bibinfo{title}{{Nuclear star clusters},} \aapr, 28, 4, \dodoi{10.1007/s00159-020-00125-0}

\bibitem[{I. Pelupessy {et~al.}(2000)Pelupessy, Lamers, \& Vink}]{2000A&A...359..695P}
Pelupessy, I., Lamers, H. J. G. L.~M., \& Vink, J.~S. 2000, The Radiation Driven Winds of Rotating {{B}}[e] Supergiants, \doarXiv{astro-ph/0005300}

\bibitem[{ {Planck Collaboration} {et~al.}(2020){Planck Collaboration}, {Aghanim}, {Akrami}, {Arroja}, {Ashdown}, {Aumont}, {Baccigalupi}, {Ballardini}, {Banday}, {Barreiro}, {Bartolo}, {Basak}, {Battye}, {Benabed}, {Bernard}, {Bersanelli}, {Bielewicz}, {Bock}, {Bond}, {Borrill}, {Bouchet}, {Boulanger}, {Bucher}, {Burigana}, {Butler}, {Calabrese}, {Cardoso}, {Carron}, {Casaponsa}, {Challinor}, {Chiang}, {Colombo}, {Combet}, {Contreras}, {Crill}, {Cuttaia}, {de Bernardis}, {de Zotti}, {Delabrouille}, {Delouis}, {D{\'e}sert}, {Di Valentino}, {Dickinson}, {Diego}, {Donzelli}, {Dor{\'e}}, {Douspis}, {Ducout}, {Dupac}, {Efstathiou}, {Elsner}, {En{\ss}lin}, {Eriksen}, {Falgarone}, {Fantaye}, {Fergusson}, {Fernandez-Cobos}, {Finelli}, {Forastieri}, {Frailis}, {Franceschi}, {Frolov}, {Galeotta}, {Galli}, {Ganga}, {G{\'e}nova-Santos}, {Gerbino}, {Ghosh}, {Gonz{\'a}lez-Nuevo}, {G{\'o}rski}, {Gratton}, {Gruppuso}, {Gudmundsson}, {Hamann}, {Handley}, {Hansen}, {Helou}, {Herranz}, {Hildebrandt}, {Hivon}, {Huang}, {Jaffe},
  {Jones}, {Karakci}, {Keih{\"a}nen}, {Keskitalo}, {Kiiveri}, {Kim}, {Kisner}, {Knox}, {Krachmalnicoff}, {Kunz}, {Kurki-Suonio}, {Lagache}, {Lamarre}, {Langer}, {Lasenby}, {Lattanzi}, {Lawrence}, {Le Jeune}, {Leahy}, {Lesgourgues}, {Levrier}, {Lewis}, {Liguori}, {Lilje}, {Lilley}, {Lindholm}, {L{\'o}pez-Caniego}, {Lubin}, {Ma}, {Mac{\'\i}as-P{\'e}rez}, {Maggio}, {Maino}, {Mandolesi}, {Mangilli}, {Marcos-Caballero}, {Maris}, {Martin}, {Martinelli}, {Mart{\'\i}nez-Gonz{\'a}lez}, {Matarrese}, {Mauri}, {McEwen}, {Meerburg}, {Meinhold}, {Melchiorri}, {Mennella}, {Migliaccio}, {Millea}, {Mitra}, {Miville-Desch{\^e}nes}, {Molinari}, {Moneti}, {Montier}, {Morgante}, {Moss}, {Mottet}, {M{\"u}nchmeyer}, {Natoli}, {N{\o}rgaard-Nielsen}, {Oxborrow}, {Pagano}, {Paoletti}, {Partridge}, {Patanchon}, {Pearson}, {Peel}, {Peiris}, {Perrotta}, {Pettorino}, {Piacentini}, {Polastri}, {Polenta}, {Puget}, {Rachen}, {Reinecke}, {Remazeilles}, {Renault}, {Renzi}, {Rocha}, {Rosset}, {Roudier}, {Rubi{\~n}o-Mart{\'\i}n},
  {Ruiz-Granados}, {Salvati}, {Sandri}, {Savelainen}, {Scott}, {Shellard}, {Shiraishi}, {Sirignano}, {Sirri}, {Spencer}, {Sunyaev}, {Suur-Uski}, {Tauber}, {Tavagnacco}, {Tenti}, {Terenzi}, {Toffolatti}, {Tomasi}, {Trombetti}, {Valiviita}, {Van Tent}, {Vibert}, {Vielva}, {Villa}, {Vittorio}, {Wandelt}, {Wehus}, {White}, {White}, {Zacchei}, \& {Zonca}}]{2020A&A...641A...1P}
{Planck Collaboration}, {Aghanim}, N., {Akrami}, Y., {et~al.} 2020, \bibinfo{title}{{Planck 2018 results. I. Overview and the cosmological legacy of Planck},} \aap, 641, A1, \dodoi{10.1051/0004-6361/201833880}

\bibitem[{S.~F. {Portegies Zwart} {et~al.}(2006){Portegies Zwart}, {Baumgardt}, {McMillan}, {Makino}, {Hut}, \& {Ebisuzaki}}]{Portegies_Zwart2006}
{Portegies Zwart}, S.~F., {Baumgardt}, H., {McMillan}, S. L.~W., {et~al.} 2006, \bibinfo{title}{{The Ecology of Star Clusters and Intermediate-Mass Black Holes in the Galactic Bulge},} \apj, 641, 319, \dodoi{10.1086/500361}

\bibitem[{S.~F. Portegies~Zwart {et~al.}(2010)Portegies~Zwart, McMillan, \& Gieles}]{Portegies_Zwart_2010}
Portegies~Zwart, S.~F., McMillan, S.~L., \& Gieles, M. 2010, \bibinfo{title}{Young Massive Star Clusters,} Annual Review of Astronomy and Astrophysics, 48, 431–493, \dodoi{10.1146/annurev-astro-081309-130834}

\bibitem[{Y. {Qin} {et~al.}(2018){Qin}, {Fragos}, {Meynet}, {Andrews}, {S{\o}rensen}, \& {Song}}]{Qin2018}
{Qin}, Y., {Fragos}, T., {Meynet}, G., {et~al.} 2018, \bibinfo{title}{{The spin of the second-born black hole in coalescing binary black holes},} \aap, 616, A28, \dodoi{10.1051/0004-6361/201832839}

\bibitem[{C. {Reisswig} {et~al.}(2009){Reisswig}, {Husa}, {Rezzolla}, {Dorband}, {Pollney}, \& {Seiler}}]{Reisswig2009}
{Reisswig}, C., {Husa}, S., {Rezzolla}, L., {et~al.} 2009, \bibinfo{title}{{Gravitational-wave detectability of equal-mass black-hole binaries with aligned spins},} \prd, 80, 124026, \dodoi{10.1103/PhysRevD.80.124026}

\bibitem[{M. Renzo {et~al.}(2020)Renzo, Farmer, Justham, G{\"o}tberg, {de Mink}, Zapartas, Marchant, \& Smith}]{renzoPredictionsHydrogenfreeEjecta2020}
Renzo, M., Farmer, R., Justham, S., {et~al.} 2020, \bibinfo{title}{Predictions for the Hydrogen-Free Ejecta of Pulsational Pair-Instability Supernovae,} Astronomy and Astrophysics, 640, A56, \dodoi{10.1051/0004-6361/202037710}

\bibitem[{M. Renzo \& N. Smith(2024)Renzo \& Smith}]{renzoPairinstabilityEvolutionExplosions2024}
Renzo, M., \& Smith, N. 2024, Pair-Instability Evolution and Explosions in Massive Stars, arXiv.
\newblock \doarXiv{2407.16113}

\bibitem[{C.~L. Rodriguez {et~al.}(2015)Rodriguez, Morscher, Pattabiraman, Chatterjee, Haster, \& Rasio}]{rodriguezBinaryBlackHole2015}
Rodriguez, C.~L., Morscher, M., Pattabiraman, B., {et~al.} 2015, \bibinfo{title}{Binary {{Black Hole Mergers}} from {{Globular Clusters}}: {{Implications}} for {{Advanced LIGO}},} Physical Review Letters, 115, 051101, \dodoi{10.1103/physrevlett.115.051101}

\bibitem[{C.~L. Rodriguez {et~al.}(2019{\natexlab{a}})Rodriguez, Zevin, {Amaro-Seoane}, Chatterjee, Kremer, Rasio, \& Ye}]{rodriguezBlackHolesNext2019}
Rodriguez, C.~L., Zevin, M., {Amaro-Seoane}, P., {et~al.} 2019{\natexlab{a}}, \bibinfo{title}{Black Holes: {{The}} next Generation---Repeated Mergers in Dense Star Clusters and Their Gravitational-Wave Properties,} Phys. Rev. D, 100, 043027, \dodoi{10.1103/PhysRevD.100.043027}

\bibitem[{C.~L. Rodriguez {et~al.}(2019{\natexlab{b}})Rodriguez, Zevin, {Amaro-Seoane}, Chatterjee, Kremer, Rasio, \& Ye}]{Rodriguez2019}
Rodriguez, C.~L., Zevin, M., {Amaro-Seoane}, P., {et~al.} 2019{\natexlab{b}}, \bibinfo{title}{Black Holes: {{The}} next Generation---Repeated Mergers in Dense Star Clusters and Their Gravitational-Wave Properties,} Physical Review D: Particles and Fields, 100, 043027, \dodoi{10.1103/PhysRevD.100.043027}

\bibitem[{I. {Romero-Shaw} {et~al.}(2020){Romero-Shaw}, {Lasky}, {Thrane}, \& {Calder{\'o}n Bustillo}}]{ecc_romeroshaw_2020}
{Romero-Shaw}, I., {Lasky}, P.~D., {Thrane}, E., \& {Calder{\'o}n Bustillo}, J. 2020, \bibinfo{title}{{GW190521: Orbital Eccentricity and Signatures of Dynamical Formation in a Binary Black Hole Merger Signal},} \apjl, 903, L5, \dodoi{10.3847/2041-8213/abbe26}

\bibitem[{T. Ryu {et~al.}(2023)Ryu, de Mink, Farmer, Pakmor, Perna, \& Springel}]{Ryu2023}
Ryu, T., de Mink, S.~E., Farmer, R., {et~al.} 2023, \bibinfo{title}{Close encounters of star–black hole binaries with single stars,} Monthly Notices of the Royal Astronomical Society, 527, 2734, \dodoi{10.1093/mnras/stad3082}

\bibitem[{S. {Sigurdsson} \& E.~S. {Phinney}(1995){Sigurdsson} \& {Phinney}}]{Sigurdsson1995}
{Sigurdsson}, S., \& {Phinney}, E.~S. 1995, \bibinfo{title}{{Dynamics and Interactions of Binaries and Neutron Stars in Globular Clusters},} Astrophysical Journal Supplement, 99, 609, \dodoi{10.1086/192199}

\bibitem[{M. Spera \& M. Mapelli(2017)Spera \& Mapelli}]{speraVeryMassiveStars2017}
Spera, M., \& Mapelli, M. 2017, \bibinfo{title}{Very Massive Stars, Pair-Instability Supernovae and Intermediate-Mass Black Holes with the Sevn Code,} Monthly Notices of the Royal Astronomical Society

\bibitem[{M. Spera {et~al.}(2015)Spera, Mapelli, \& Bressan}]{Spera2015}
Spera, M., Mapelli, M., \& Bressan, A. 2015, \bibinfo{title}{The Mass Spectrum of Compact Remnants from the {{PARSEC}} Stellar Evolution Tracks,} 451, 4086, \dodoi{10.1093/mnras/stv1161}

\bibitem[{M. Spera {et~al.}(2019)Spera, Mapelli, Giacobbo, Trani, Bressan, \& Costa}]{speraMergingBlackHole2019a}
Spera, M., Mapelli, M., Giacobbo, N., {et~al.} 2019, \bibinfo{title}{Merging Black Hole Binaries with the {{SEVN}} Code,} Monthly Notices of the Royal Astronomical Society, 485, 889, \dodoi{10.1093/mnras/stz359}

\bibitem[{T. Sukhbold {et~al.}(2016)Sukhbold, Ertl, Woosley, Brown, \& Janka}]{sukhbold_core-collapse_2016}
Sukhbold, T., Ertl, T., Woosley, S.~E., Brown, J.~M., \& Janka, H.-T. 2016, \bibinfo{title}{{{CORE-COLLAPSE SUPERNOVAE FROM}} 9 {{TO}} 120 {{SOLAR MASSES BASED ON NEUTRINO-POWERED EXPLOSIONS}},} The Astrophysical Journal Letters, 821, 38, \dodoi{10.3847/0004-637X/821/1/38}

\bibitem[{H. {Tagawa} {et~al.}(2020){Tagawa}, {Haiman}, \& {Kocsis}}]{2020ApJ...898...25T}
{Tagawa}, H., {Haiman}, Z., \& {Kocsis}, B. 2020, \bibinfo{title}{{Formation and Evolution of Compact-object Binaries in AGN Disks},} \apj, 898, 25, \dodoi{10.3847/1538-4357/ab9b8c}

\bibitem[{ {The LIGO Scientific Collaboration} {et~al.}(2025){The LIGO Scientific Collaboration}, {the Virgo Collaboration}, \& {the KAGRA Collaboration}}]{GW231123_paper}
{The LIGO Scientific Collaboration}, {the Virgo Collaboration}, \& {the KAGRA Collaboration}. 2025, \bibinfo{title}{{GW231123: a Binary Black Hole Merger with Total Mass 190-265 $M_{\odot}$},} arXiv e-prints, arXiv:2507.08219, \dodoi{10.48550/arXiv.2507.08219}

\bibitem[{S. {Torniamenti} {et~al.}(2024){Torniamenti}, {Mapelli}, {P{\'e}rigois}, {Arca Sedda}, {Artale}, {Dall'Amico}, \& {Vaccaro}}]{Torniamenti2024}
{Torniamenti}, S., {Mapelli}, M., {P{\'e}rigois}, C., {et~al.} 2024, \bibinfo{title}{{Hierarchical binary black hole mergers in globular clusters: Mass function and evolution with redshift},} \aap, 688, A148, \dodoi{10.1051/0004-6361/202449272}

\bibitem[{C. {Ugolini} {et~al.}(2025){Ugolini}, {Limongi}, {Schneider}, {Chieffi}, {Di Carlo}, \& {Spera}}]{Ugolini2025}
{Ugolini}, C., {Limongi}, M., {Schneider}, R., {et~al.} 2025, \bibinfo{title}{{The initial mass-remnant mass relation for core collapse supernovae},} arXiv e-prints, arXiv:2501.18689, \dodoi{10.48550/arXiv.2501.18689}

\bibitem[{L. {van Son} {et~al.}(2021){van Son}, Justham, \& De~Mink}]{vansonFillingGapPossible2021}
{van Son}, L., Justham, S., \& De~Mink, S. 2021, \bibinfo{title}{Filling the Gap: Possible Pathways to Get a {{BBH}} Merger in the {{PISN}} Mass Gap,} 43, 2124

\bibitem[{J.~S. Vink {et~al.}(2018)Vink, {de Koter}, \& Lamers}]{vinkPresupernovaEvolutionExplosive2018}
Vink, J.~S., {de Koter}, A., \& Lamers, H. J. G. L.~M. 2018, \bibinfo{title}{Presupernova {{Evolution}} and {{Explosive Nucleosynthesis}} of {{Rotating Massive Stars}} in the {{Metallicity Range}} -3 {$\leq$} [{{Fe}}/{{H}}] {$\leq$} 0,} The Astrophysical Journal Supplement Series, 237, 13, \dodoi{10.3847/1538-4365/aacb24}

\bibitem[{N.~C. {Weatherford} {et~al.}(2021){Weatherford}, {Fragione}, {Kremer}, {Chatterjee}, {Ye}, {Rodriguez}, \& {Rasio}}]{weatherford_2021}
{Weatherford}, N.~C., {Fragione}, G., {Kremer}, K., {et~al.} 2021, \bibinfo{title}{{Black Hole Mergers from Star Clusters with Top-heavy Initial Mass Functions},} \apjl, 907, L25, \dodoi{10.3847/2041-8213/abd79c}

\bibitem[{S.~E. Woosley(2017{\natexlab{a}})Woosley}]{Woosley2017}
Woosley, S.~E. 2017{\natexlab{a}}, \bibinfo{title}{Pulsational Pair-Instability Supernovae,} 836, 244, \dodoi{10.3847/1538-4357/836/2/244}

\bibitem[{S.~E. Woosley(2017{\natexlab{b}})Woosley}]{woosleyPulsationalPairinstabilitySupernovae2017}
Woosley, S.~E. 2017{\natexlab{b}}, \bibinfo{title}{Pulsational {{Pair-instability Supernovae}},} The Astrophysical Journal, 836, 244, \dodoi{10.3847/1538-4357/836/2/244}

\bibitem[{M. Zevin {et~al.}(2021)Zevin, Bavera, Berry, Kalogera, Fragos, Marchant, Rodriguez, Antonini, Holz, \& Pankow}]{zevinOneChannelRule2021a}
Zevin, M., Bavera, S.~S., Berry, C. P.~L., {et~al.} 2021, \bibinfo{title}{One {{Channel}} to {{Rule Them All}}? {{Constraining}} the {{Origins}} of {{Binary Black Holes Using Multiple Formation Pathways}},} The Astrophysical Journal, 910, 152, \dodoi{10.3847/1538-4357/abe40e}

\bibitem[{R. {Zinn} \& M.~J. {West}(1984){Zinn} \& {West}}]{1984ApJS...55...45Z}
{Zinn}, R., \& {West}, M.~J. 1984, \bibinfo{title}{{The globular cluster system of the Galaxy. III. Measurements of radial velocity and metallicity for 60 clusters and a compilation of metallicities for 121 clusters.},} \apjs, 55, 45, \dodoi{10.1086/190947}

\end{thebibliography}
\bibliographystyle{aasjournalv7}

\end{document}